\documentclass[12pt,a4paper]{article}
\usepackage{amsmath,amssymb,mathrsfs,framed,esint,slashed,upgreek}
\usepackage[colorlinks]{hyperref}
\usepackage{color}
\usepackage[all]{xy}

\usepackage{soul}

\newlength{\bibitemsep}
\newlength{\bibparskip}
\let\oldthebibliography\thebibliography
\renewcommand\thebibliography[1]{%
  \oldthebibliography{#1}%
  \setlength{\parskip}{\bibitemsep}%
  \setlength{\itemsep}{\bibparskip}%
}

\DeclareMathAlphabet{\mathpzc}{OT1}{pzc}{m}{it}

\newcommand{\rem}[1]{}
\newcommand{\bLambda}{{\boldsymbol{\Lambda}}}

\newcommand{\de}{{\rm d}}

\newcommand{\bv}{{\mathbf{v}}}

\newcommand{\bX}{{\mathbf{X}}}

\newcommand{\bzeta}{{\boldsymbol{\zeta}}}
\newcommand{\bsigma}{{\boldsymbol{\sigma}}}
\newcommand{\bnu}{{\boldsymbol{\nu}}}

\newcommand{\blambda}{{\boldsymbol{\lambda}}}
\newcommand{\bz}{{\mathbf{{\boldsymbol{z}}}}}
\newcommand{\bw}{{\mathbf{w}}}

\newcommand{\bxi}{{\boldsymbol{\xi}}}

\newcommand{\bgamma}{\boldsymbol{\gamma}}
\newcommand{\beq}{\begin{equation}}
\newcommand{\eeq}{\end{equation}}
\newcommand{\ben}{\begin{eqnarray}}
\newcommand{\een}{\end{eqnarray}}

\renewcommand{\contentsname}{}

\numberwithin{equation}{section}

\def\contract{\makebox[1.2em][c]{\mbox{\rule{.6em}
{.01truein}\rule{.01truein}{.6em}}}}

\begin{document}

\title{\vspace{-1cm}Evolution of hybrid quantum-classical wavefunctions
\vspace{-.3cm}
}
\author{Fran\c{c}ois Gay-Balmaz$^1$, Cesare Tronci$^{2,3}$ \smallskip\vspace{-.1cm}
\\ \vspace{-.1cm}
\footnotesize
\it $^1$CNRS and \'Ecole Normale Sup\'erieure, Laboratoire de M\'et\'eorologie Dynamique, Paris, France
\\\vspace{-.1cm}
\footnotesize
\it $^2$Department of Mathematics, University of Surrey, Guildford, United Kingdom
\\\vspace{-.1cm}
\footnotesize
\it 
$^3$Department of Physics and Engineering Physics, Tulane University, New Orleans LA, United States\vspace{-.5cm}}
\date{}

\maketitle

\begin{abstract}\footnotesize
A  gauge-invariant wave equation for the dynamics of hybrid quantum-classical systems is formulated by combining the variational setting of Lagrangian paths in continuum theories with  Koopman wavefunctions in classical mechanics. We identify gauge transformations with unobservable phase factors in the classical phase-space and we introduce gauge invariance in the  variational principle underlying a  hybrid wave equation previously proposed by the authors. While the original construction ensures a positive-definite quantum density matrix,  the present  model also guarantees the same property  for the classical Liouville density. After a suitable wavefunction factorization,  gauge invariance is achieved by resorting to the classical Lagrangian paths made available by the Madelung transform of  Koopman wavefunctions.  Due to the appearance of a phase-space analogue of the Berry connection, the new hybrid wave equation is highly nonlinear and it is proposed here  as a platform for further developments in  quantum-classical dynamics. Indeed, the associated model is Hamiltonian and appears to be the first to ensure a series of consistency properties beyond positivity of quantum and classical densities. For example, the model  possesses a quantum-classical Poincar\'e integral invariant and its special cases  include both the mean-field model and the Ehrenfest model from chemical physics. 
\end{abstract}

\vspace{-1.3cm}
{
\contentsname
\footnotesize
\tableofcontents
}
\addtocontents{toc}{\protect\setcounter{tocdepth}{3}}

\section{Introduction}

The coupling between classical and quantum degrees of freedoms represents an outstanding question ever since the rise of quantum mechanics as a physical theory. Following Bohr's work,  Landau and Lifshitz \cite{LaLi77} defined a \emph{quantum measurement} as the interaction between a measured quantum system and a classical system commonly referred to as the \emph{apparatus}. In this way, ``quantum mechanics occupies a very unusual place among physical theories: it contains classical mechanics as a limiting case, yet at the same time it requires this limiting case for its own formulation'' \cite{LaLi77}. However, the search for an extension of quantum theory (and its mathematical footing) to include the coupling to a classical system has been one of the most challenging directions for about a century. The main difficulty is capturing the correlations between the quantum system and the `apparatus' that are responsible for the \emph{decoherence} phenomenon, i.e. the loss of pure states in the quantum sector as time goes by. In addition, according to Bohr \cite{Bo35}, the classical apparatus itself undergoes some kind of uncontrollable disturbance that is also a consequence of quantum-classical correlations. While Bohr's disturbance has never been fully characterized, a possible interpretation lies in the loss of classical pure states, i.e. the loss of delta-like classical Liouville distributions \cite{BoGBTr19}. 

The search for an extension of quantum theory to incorporate the interaction with classical systems leads naturally to the idea of a hybrid model in which classical and quantum dynamics are treated on an equal footing: when no interaction is present in the treatment, the theory must reduce to the uncoupled equations of quantum and classical mechanics, that is Schr\"odinger's (or quantum Liouville) and Newton's (or classical Liouville) equations, respectively. For example, this important property is satisfied by the mean-field model.
In standard Poisson bracket and commutator notation, this model is given by
\beq
\frac{\partial \rho_c}{\partial t}=\{\operatorname{\sf Tr} (\hat{\rho}_q\widehat{H}),\rho_c \}
\,,\qquad \qquad
i\hbar\frac{\partial {\hat\rho}_q}{\partial t}=\bigg[\int\!\rho_c\widehat{H}\de q\de p\,,\hat{\rho}_q\bigg]
.
\label{MFeqs}
\eeq
Here, $\rho_c$ and ${\hat\rho}_q$ are the classical Liouville density and the quantum von Neumann operator, while the hybrid Hamiltonian $\widehat{H}(q,p)$ is a phase-space function taking values in the quantum operators. Unfortunately, the mean-field equations \eqref{MFeqs} neglect quantum-classical correlations \cite{FaJiSp18}, thereby ignoring decoherence. Thus, one is motivated to look for a more involved model where the fundamental quantities are functions of both classical and quantum coordinates on phase-space and configuration-space, respectively. This direction was pursued only occasionally in the physics community \cite{boucher,PrKi,Sudarshan,WiSr92} and with little success. For example, in some cases the models fail to reduce to the uncoupled equations of quantum and classical mechanics in the absence of quantum-classical interaction \cite{Hall}. {\color{black}Lists of consistency criteria for hybrid quantum-classical models were provided in \cite{boucher} and more recently in \cite{GBTr21}. 
Among the most important criteria, we find the positivity of quantum and classical densities, as well as the decoherence property $\operatorname{\sf Tr}{\hat\rho}_q^2\neq const.$}

The difficulties encountered in establishing a hybrid quantum-classical theory beyond mean-field led much of the community to move away from this direction, thereby abandoning Bohr's initial view of quantum measurements based on  quantum-classical interaction. For example, no-go arguments have been proposed over the years to even exclude the possibility of such hybrid theories \cite{Marletto,Salcedo,Terno} {\color{black}beyond  simple models of mean-field type}. A currently common picture involves a fully quantum apparatus in thermodynamic equilibrium, while the link to classical mechanics is provided by the classical limit that is obtained as a result of long-time evolution. Based on Lindblad's master equation \cite{Li76} this approach completely ignores the motion of the apparatus, whose degrees of freedom are traced out to yield the irreversible dynamics of the quantum density matrix. Due to irreversibility, the fine small-scale structures are eventually smoothened out just as in the classical Fokker-Planck equation. Then, in this picture a classical state emerges as the result of irreversible dissipative dynamics at the quantum scales. Due to the absence of small-scale phenomenology, it has been pointed out that the classical states made available in this way are ``arguably among the least classical ones'' \cite{ScCa11}, as it seems to emerge also from recent simulations in the Wigner picture \cite{CaBoJaRa15}. However, we shall not enter this discussion here. 

While quantum-classical coupling first emerges as a foundational question in relation to the measurement process, similar question appear in semiclassical descriptions of gravity \cite{AlKiRe08}. In this context, the absence of a consistent quantum theory of gravity leads to treating the geometric background as a classical system \cite{BoDi21}. Then, the question is how an evolving quantum field in an expanding universe affects the expansion rate of the background on which it is defined. Currently available semiclassical theories fail to address this \emph{backreaction problem} and thus one is motivated to look for more complete  models that can include this important effect \cite{boucher}.

Besides the foundational aspects, the development of mixed quantum-classical models is also an active area of research in more applied areas such as chemical physics \cite{CrBa18,Tavernelli,Tully}, solid state physics \cite{BACaCaVe09,HuHeMa17}, and more recently spintronics \cite{RuKaUp22}. For example, in the latter  case, recent proposals involve the use of classical ferromagnets to enlarge the parameter space in spin control. On the other hand, in quantum chemistry and solid-state physics, hybrid/mixed quantum-classical models emerge as a promising direction to alleviate the curse of dimensionality currently challenging many-particle quantum simulations. One typically identifies the slow motion of some degrees of freedom which are eventually treated as classical in order to mitigate the cost of simulating their evolution. The most celebrated example is provided by  Born-Oppenheimer molecular dynamics  \cite{BoOp27}, which arises from methods in slow-manifold reduction \cite{MacKay}. In this model, the nuclei are treated as classical, while quantum electrons are constrained to evolve adiabatically in their ground state. Then, the degeneracies appearing in the associated electronic eigenvalue problem lead to topological singularities, known as ``conical intersections''. The numerical treatment of these singularities and their resulting geometric phase pose further difficulties that continue to challenge the community. However, recently the question emerged whether the topological singular nature should be retained at all in order to account for geometric phase effects \cite{RePrGr17,RaTr20}. 

Despite the paramount importance of Born-Oppenheimer molecular dynamics, the adiabatic approximation underlying electron dynamics is quite limitative and a whole new industry has now emerged to tackle nonadiabatic dynamics. In this context, the equations currently implemented in hybrid quantum-classical methods \cite{CrBa18} often lack a sound mathematical footing. The desire for such a fundamental rigorous ground has led some \cite{Ka16,SuOuLa13} to derive mixed quantum-classical algorithms from the Aleksandrov-Gerasimenko (AG) equation \cite{Aleksandrov,Gerasimenko}
\beq\label{AGeq}
\frac{\partial\widehat{\cal D}}{\partial t}=-i\hbar^{-1}[\widehat{H},\widehat{\cal D}]+\frac12\left(\{\widehat{H},\widehat{\cal D}\}-\{\widehat{\cal D},\widehat{H}\}\right).
\eeq
Here, $\widehat{\cal D}(q,p)$ is a phase-space density with values in the quantum von Neumann operators, so that the classical and quantum densities are written as $\rho_c=\operatorname{\sf Tr}\widehat{\cal D}$ and $\hat{\rho}_q=\int\!\widehat{\cal D}\,\de q\de p$.  
Equation \eqref{AGeq} satisfies several important properties \cite{boucher} and is commonly regarded as a fundamental step beyond the mean-field model. In turn, the AG equation allows for a sign-indefinite density matrix $\hat{\rho}_q$ of the quantum subsystem with the potential drawback of violating the uncertainty principle. In addition, as the  AG equation breaks the time-reversal symmetry, one would like to justify this property in terms of an $H-$theorem for the entropy increase. However, such a result is currently unavailable. This lack of entropy arguments also applies to stochastic augmentations of the AG equation \cite{Diosi}.

Recently, a new approach has been proposed by the authors \cite{BoGBTr19} upon following a suggestion by George Sudarshan \cite{Marmo,Sudarshan}. By exploiting Koopman's Hilbert space formulation of classical mechanics \cite{Koopman}, one can try to construct a hybrid quantum-classical wavefunction  $\Upsilon(q,p,x)$, whose dynamics recovers the mean-field model by writing $\Upsilon(q,p,x)=\chi(q,p)\psi(x)$. Here, $(q,p)$ are classical phase-space coordinates and $x$ is the quantum configuration coordinate. Moreover, $\chi$ is a Koopman wavefunction obeying an equation of the type 
\beq\label{KvN+gauge}
i\hbar\partial_t\chi=\{i\hbar H,\chi\}+\varphi\chi
\,,
\eeq
so that the classical Liouville equation $\partial_t\rho_c=\{H,\rho_c\}$ emerges by writing $\rho_c(q,p)=|\chi(q,p)|^2$.  However, finding a consistent evolution law for  a general hybrid wavefunction $\Upsilon(q,p,x)$ is far from easy and Sudarshan's early attempt led to interpretative issues \cite{Barcelo,PeTe,Terno}. The authors recently addressed this problem by modifying Koopman's original prescription $\rho_c=|\chi|^2$ after selecting the specific phase factor $\varphi=p\partial_pH-H$  given by  the usual Lagrangian on phase-space. The resulting hybrid theory leads to the construction of an unsigned operator-valued distribution $\widehat{\cal D}(q,p)$, which however yields a positive-definite quantum density matrix $\hat{\rho}_q$. As this feature allows to recover the quantum uncertainty principle, the underlying hybrid model from \cite{BoGBTr19} represents a step forward beyond the AG equation \eqref{AGeq}. 
However, the classical density $\rho_c$ constructed in this way has a sign-indefinite expression and thus one is led to ask whether an initially positive Liouville density may develop negative values in time.  As discussed in \cite{GBTr20}, a general answer is currently unavailable, although there are infinite families of hybrid systems for which $\rho_c$ was shown to stay positive in time.

The question whether the classical density should be  positive-definite in hybrid dynamics is crucial since the development of negative values may lead to interpretative issues. Some of these issues were addressed in \cite{BoGBTr19} by resorting to analogies with Wigner functions, following earlier ideas by Feynman \cite{Feynman}. However, general arguments are still lacking and thus one may  ask if the hybrid theory presented in \cite{BoGBTr19} can be modified to accommodate a positive-definite classical Liouville density. A positive answer was provided by the authors in \cite{GBTr21} by combining recent mathematical techniques in chemical physics and the geometry of Hamilton's variational principle. The main feature is that, while the original hybrid theory is linear, the new closure variant appears to be nonlinear and the nonlinearity seems to be produced by the emergence of a Berry connection in phase-space. {\color{black}Explicitly, this closure module reads as follows:
\beq
i\hbar\partial_t \widehat{\mathcal{P}}+i\hbar\operatorname{div} \!\big( \widehat{\mathcal{P}}\left\langle \bX_{ \widehat{\cal H}}\right\rangle\!\big)=\big[\widehat{\cal H},\widehat{\mathcal{P}}\big]\,,
\label{andrea}
\eeq
where $\widehat{\mathcal{P}}(q,p)$ is a hybrid density playing the role of $\widehat{\mathcal{D}}$ in \eqref{AGeq}, while we have used the notation $\bX_{\widehat{A}}=(\partial_p\widehat{A},-\partial_q\widehat{A}\,)$ and $\langle\widehat{A}\rangle=\operatorname{\sf Tr}(\widehat{A}\widehat{\mathcal{P}})/\operatorname{\sf Tr}\widehat{\mathcal{P}}$. Here, the effective Hamiltonian $\widehat{\cal H}$ is given by $\widehat{\cal H}=\widehat{H}+\hbar\widehat{\cal F}(\widehat{\cal P})$, where the operator-valued function $\widehat{\cal F}$ is expressed in terms of $\widehat{\cal P}$ and $\{\widehat{\cal P},\widehat{H}\}+\{\widehat{H},\widehat{\cal P}\}$. Neglecting the $\hbar-$contribution to the effective Hamiltonian reduces the equation \eqref{andrea} to the \emph{Ehrenfest model}, which will be discussed at the end of this paper. In more generality, the full closure model \eqref{andrea} has so far been formulated only in terms of hybrid quantum-classical densities. In this paper, we will show that a wavefunction description unfolds several relevant properties, including the existence of a Poincar\'e integral invariant leading to the construction of a Liouville volume density that can be used to identify a Gibbs entropy functional for the classical sector.}

In particular, this paper shows that positivity of the classical density may be achieved by adopting a gauge invariance principle to reflect the fact that classical phases are actually unobservable. This particular remark about the irrelevance of local phases in classical mechanics represents the second fundamental point raised by Sudarshan and it was addressed in his paper by the enforcement of specific superselection rules whose mode of operation has never been clear \cite{Barcelo,PeTe,Terno}. Instead of superselection rules, this paper addresses the point of unobservable classical phases by treating them as a gauge freedom. This is exactly what happens in standard quantum mechanics where global phase factors are irrelevant and wavefunctions are always defined up to a phase. In this context, the Liouville density can be thought of as a Noether charge just like total probability in quantum theory. This paper pursues this analogy by making the original treatment in \cite{BoGBTr19} manifestly gauge invariant under local phase transformations, that is multiplicative phase factors for which the phase is a function on the classical phase-space. The achievement of  gauge invariance is made possible by the specific fact that the original quantum-classical wave equation in \cite{BoGBTr19} is already manifestly gauge-covariant.
On the other hand, the actual process leading to gauge invariance requires several steps. Starting with the variational principle underlying the original theory, its associated Lagrangian will undergo several exact transformation steps so that it is finally taken into a form where the gauge invariance principle can finally be applied. As we will see, this is a nontrivial process combining hydrodynamic approaches to Liouville phase-space dynamics with specific wavefunction factorization techniques recently developed in chemical physics. The combination of these approaches is made necessary by two facts: 1) the classical Liouville equation is a characteristic equation accompanied by a Lagrangian-path hydrodynamic description, and 2) the extraction of the classical phase in a hybrid context requires writing the quantum-classical wavefunction in a convenient way so that a classical phase can be adequately identified. Fortunately, the process leading to this gauge-invariance principle has a purely classical analogue, which will serve as a pilot example. As a general methodology, we emphasize that this paper   will not proceed by operating on a particular set of equations of motion. Instead, we will mostly manipulate its underlying variational principle in order to keep track of basic conservation laws and transformation properties. This is a common method in gauge theory and has a long history in quantum dynamics \cite{KrSa81}.

\medskip
\noindent{\bf Plan of the paper.} Our discussion starts with a discussion of Koopman wavefunctions in classical mechanics. After a preliminary introduction to standard  Koopman-von Neumann (KvN) theory in Section \ref{sec:KvN}, we show how apparent issues in the original treatment may be addressed by adding a phase term first proposed by van Hove. This is presented in Section \ref{sec:KvH}, which covers the Koopman-van Hove (KvH) formulation of classical mechanics. The important role of phase transformations in the Koopman context is discussed in Section \ref{sec:covariance}, where the concept of gauge transformation in classical mechanics is also introduced. As the KvH construction appears to differ substantially from KvN theory, one is led to ask whether these two formulations are actually related. As shown in Section \ref{sec:BackToKvN},  standard KvN dynamics may be obtained by simply applying a gauge-invariance principle to the variational formulation of KvH theory. As shown in Section \ref{sec:EPMad}, the application of this gauge-invariance principle requires techniques from continuum theories, such as the Lagrangian flow paths widely used in hydrodynamics. Then, the gauge-invariance principle leads to an alternative variational formulation of KvN theory, which represents the immediate extension of the well-known variational setting of the classical Liouville equation \cite{TrJo21}. 

After discussing the various Koopman formulations of classical mechanics, the paper proceeds in Section \ref{sec:hybrids} to illustrate the present status of the Koopman hybrid theory for the description of mixed quantum-cassical systems \cite{BoGBTr19}. Once hybrid quantum-classical wavefunctions are introduced in Section \ref{sec:hybWF}, Section \ref{sec:hybden1} discusses how classical and quantum densities may be obtained in this context. The rest of the paper proceeds to apply the same gauge-invariance principle from Koopman classical mechanics in the hybrid setting. This requires several steps, the first of which consists in a suitable  factorization of the hybrid wavefunction, as discussed in Section \ref{sec:EF}. The latter is entirely devoted to transforming the variational principle of the original hybrid formulation from the preceding sections in such a way that the gauge principle can finally be applied. After applying the Madelung transform in Section \ref{sec:HydFrame}, once again we observe the emergence of Lagrangian flow paths in the Koopman context, as discussed in Section \ref{sec:QWFCFP}.  

At this point, the variational principle of the original hybrid formulation is amenable to the application of the gauge principle in such a way to make classical phases unobservable. This is the topic of Section \ref{sec:NHWE}, which starts by constructing the gauge-invariant Lagrangian of quantum-classical dynamics. {\color{black}The latter is formuated in Section \ref{sec:HybLagrFun} while Section \ref{sec:NHWEeq} presents the full gauge-invariant evolution of hybrid wavefunctions}.

A discussion of some of the various properties of the new gauge-invariant model is presented in Section \ref{sec:discussion}. After discussing the presence of a Poincar\'e integral invariant and some of its consequences in Section \ref{sec:PIinv}, the noncanonical Hamiltonian structure of hybrid dynamics is presented in Section \ref{sec:GIHS}. Also, Section \ref{sec:HybDenOp} presents a prototype for a quantum-classical density operator: while this appears to be generally unsigned, it gives rise to positive quantum and classical densities and possesses important covariance properties finally ensuring basic conservation laws such as momentum conservation. Finally, Section \ref{sec:SpecsFC} presents relevant special cases, including the Ehrenfest model from chemical physics. As discussed therein, the present model allows to identify explicitly the source of backreaction forces.

\section{Koopman wavefunctions in classical mechanics}

This section reviews different formulations of classical mechanics based on Koopman wavefunctions. As we will see, the standard Koopman-von Neumann formulation presents important issues which will be addressed by exploiting the variational structure of Koopman dynamics. 

\subsection{Koopman-von Neumann construction\label{sec:KvN}}
The simplest example of Koopman classical dynamics is given by the Koopman-von Neumann (KvN) equation in the following common form:
\beq
i\hbar\partial_t\chi=\{i\hbar H,\chi\}=:\widehat{L}_H\chi
\,.
\label{KvN1}
\eeq
Here, the factor $i\hbar$ is inserted to produce the Hermitian operator $\widehat{L}_H=\{i\hbar H,\,\}$, known as {\it Liouvillian}. In this way, classical dynamics is regarded as a unitary flow on the Hilbert space $\mathscr{H}_C=L^2(\Bbb{R}^2)$ of square-integrable functions on phase-space. Here, we focus on one-dimensional degrees of freedom as the extension to higher dimensions is straightforward. In polar form $\chi=\sqrt{D}e^{iS/\hbar}$, one finds
\beq
\partial_tD=\{ H,D\}
\,,\qquad\qquad
\partial_tS=\{ H,S\}
\,,
\label{KvN2}
\eeq
thereby recovering the usual Liouville equation for the density $D=|\chi|^2$. We observe that the phase can be simply set to zero so that the KvN wavefunction may be identified with a real-valued amplitude  $|\chi|$.

Since the KvN equation \eqref{KvN1} identifies classical dynamics  as a type of unitary flow, one may look at its underlying variational principle. Mimicking the Dirac-Frenkel Lagrangian for the standard Schr\"odinger equation \cite{Fr34}, we write \cite{Do14,Jo20}
\beq
\delta\int^{t_2}_{t_1}\!L_{\rm KvN}(\chi,\partial_t\chi)\,\de t=0
\,,\qquad\quad
L_{\rm KvN}(\chi,\partial_t\chi)=\operatorname{Re}\!\int\!i\hbar\chi^*(\partial_t\chi-\{H,\chi\})\,\de q\de p
\,.
\label{VP-KvN1}
\eeq
In \cite{BoGBTr19}, we showed how this variational principle for KvN dynamics may lead to relevant issues. For example, integration by parts leads to rewriting the conserved Hamiltonian functional $h_{\rm KvN} ( \chi ) =\operatorname{Re}\!\int\!i\hbar\chi^*\{H,\chi\}\,\de q\de p$ as
\beq
h_{\rm KvN}(\chi)=\hbar\int\! H\operatorname{Im}\{\chi^*,\chi\}\,\de q\de p
\,,
\label{KvNHam}
\eeq
which differs from the physical energy $\int\! H|\chi|^2\,\de q\de p$, unless one sets $|\chi|^2=\hbar\operatorname{Im}\{\chi^*,\chi\}$. Indeed, both  quantities $|\chi|^2$ and $\hbar\operatorname{Im}\{\chi^*,\chi\}$ satisfy the classical Liouville equation, although they can  be set  equal only by allowing for topological singularities in the KvN wavefunction \cite{Jo20}.
More importantly,  if one initializes a zero phase $S=0$ in \eqref{KvN2}, then the  Hamiltonian functional \eqref{KvNHam} collapses entirely, thereby indicating that this variational formulation may need special care. This difficulty was recently overcome in \cite{TrJo21}, where an alternative variational formulation was presented. This  variant will be discussed later on. Instead, the next section shows how this and other apparent issues may be addressed by modifying the original KvN formulation.

\subsection{Koopman-van Hove construction\label{sec:KvH}}

One possibility to overcome the difficulties arising from the conventional variational formulation of KvN theory was presented in \cite{BoGBTr19} and follows from previous work by van Hove \cite{VanHove} and Kostant \cite{Kostant} on prequantum theory in geometric quantization \cite{Fa07,Souriau}. Without going into the details, the immediate modification consists in changing the phase dynamics of KvN theory by replacing  \eqref{KvN2} by
\beq
\partial_tD=\{ H,D\}
\,,\qquad\qquad 
\partial_tS+\{ S,H\}=p\partial_pH-H=:\mathscr{L}
\,,
\label{KvH2Mad}
\eeq
where $\mathscr{L}$ identifies the usual phase-space expression of the Lagrangian which already appeared in the Introduction. 
Notice that the same equations may be written in the characteristic form
\[
\frac{\de}{\de t}D(q,p,t)=0
\qquad\ 
\frac{\de}{\de t}S(q,p,t)=\mathscr{L}(q,p)
\,,\qquad \text{along}\qquad
(\dot{q},\dot{p})=(\partial_pH,-\partial_qH)=:\bX_H(q,p)
\,,
\]
where we have used standard notation for the Hamiltonian vector field $\bX_H(q,p)$. We notice that the phase equation is reminiscent of the usual  phase dynamics occurring in Hamilton-Jacobi theory;  see \cite{deGo04} for more details on this point. 

The equations \eqref{KvH2Mad} are naturally written in terms of the wavefunction $\chi=\sqrt{D}e^{iS/\hbar}$, thereby leading to the {\it Koopman-van Hove} (KvH) equation
\beq
i\hbar
\partial_t\chi=i\hbar\{ H,\chi\}-\mathscr{L}\chi=:\widehat{\cal L}_H\chi
\,.
\label{KvH2}
\eeq
Following \cite{BoGBTr19}, here we have introduced the \emph{covariant Liouvillian operator} $\widehat{\cal L}_H:=i\hbar\{ H,\ \}-\mathscr{L}$, which is also known in prequantum theory as \emph{prequantum operator} \cite{Kostant, Souriau}. This Hermitian operator replaces the standard Liouvillian  $\widehat{L}_H:=i\hbar\{ H,\ \}$ appearing in the KvN equation \eqref{KvN1}. The sense in which $\widehat{\cal L}_H$ is covariant was explained in \cite{BoGBTr19} in terms of minimal coupling arguments in phase-space. See Section \ref{sec:covariance} for further comments on gauge covariance in KvH theory.
As equation \eqref{KvH2} again identifies a unitary flow on  $\mathscr{H}_C=L^2(\Bbb{R}^2)$ analogous to the Schr\"odinger equation, one can again look at the underlying variational principle of Dirac-Frenkel type, that is
\beq
\delta\int^{t_2}_{t_1}\!L_{\rm KvH}(\chi,\partial_t\chi)\,\de t=0
\,,\qquad\quad
L_{\rm KvH}(\chi,\partial_t\chi)=\operatorname{Re}\!\int\!\chi^*\Big(i\hbar\partial_t\chi-i\hbar\{H,\chi\}+\chi\mathscr{L}\Big)\de q\de p
\,.
\label{VP-KvH}
\eeq
We observe that, upon expanding $\mathscr{L}=p\partial_pH-H$ and integrating by parts, this variational principle identifies the conserved Hamiltonian functional
\beq
h_{\rm KvH}(\chi)=\int \!H\Big(|\chi|^2-\partial_p(p|\chi|^2)+\hbar\operatorname{Im}\{\chi^*,\chi\}\Big)\,\de q\de p
\,.
\label{KvHHam}
\eeq
As pointed out in \cite{BoGBTr19}, one notes that in the parenthesis both the first term and the sum of the last two  obey the classical Liouville equation. This observation leads to defining the Liouville density in KvH theory as 
\begin{align}
\rho_c=&\ |\chi|^2-\partial_p(p|\chi|^2)+\hbar\operatorname{Im}\{\chi^*,\chi\}
\nonumber
\\
\label{KvHmomap}
=&\, D-\partial_p(pD)+\{D,S\}
\,,
\end{align}
which eliminates all the ambiguities otherwise occurring in the KvN construction. For example, as shown in \eqref{KvH2}, the KvH phase undergoes nontrivial dynamics and therefore the Hamiltonian functional \eqref{KvHHam}  cannot be made to vanish by simply setting an initial zero phase. In addition, we notice that the distribution \eqref{KvHmomap} appearing in \eqref{KvHHam} integrates to 1, unlike the expression $\hbar\operatorname{Im}\{\chi^*,\chi\}$ appearing in \eqref{KvNHam}.

While the present discussion omits the mathematical details, we point out that the above expression has a deep geometric structure in terms of momentum maps, {\color{black}as discussed in}  \cite{GBTr20}. Importantly, here we notice that the expression \eqref{KvHmomap} of the classical density is not positive-definite. This point should not be regarded as an issue in classical dynamics: since the Liouvillle equation is a characteristic equation, the sign of the initial condition is preserved in time. For example, following \cite{BoGBTr19}, one can initialize the KvH wavefunction $\chi$ corresponding to a Gaussian density.

\subsection{Phase transformations in Koopman mechanics\label{sec:covariance}}
A distinctive difference between the KvN and KvH formalisms lies in the different transformation properties underlying the classical Liouville density $\rho_c$ under local $U(1)$ transformations. Indeed, while the KvN prescription $\rho_c=|\chi|^2$ is evidently invariant under $\chi\mapsto\chi e^{i\varphi/\hbar}$, the KvH expression \eqref{KvHmomap} involves a phase shift $S\mapsto S + \varphi$. Nevertheless, as pointed out in \cite{BoGBTr19}, the KvH prescription still enjoys a covariance property. To see this, let us introduce the symplectic potential $\boldsymbol{\cal A}:=(p,0)$ (equivalently,  $\boldsymbol{\cal A}(\bz)\cdot\de\bz=p\de q$) to rewrite \eqref{KvHmomap} as
\beq
\rho_c=D+\operatorname{div}(DJ(\nabla S-\boldsymbol{\cal A}))
\,,\qquad\text{with}\qquad
J=\left(\begin{matrix}0 & 1 \\ -1 & 0\end{matrix}\right),
\label{giggia}
\eeq
where $J=(\nabla\boldsymbol{\cal A})^T-\nabla\boldsymbol{\cal A}$ is the canonical symplectic form. Then, it is clear that combining the phase transformation $\chi\mapsto\chi e^{i\varphi/\hbar}$ with the shift $\boldsymbol{\cal A}\to\boldsymbol{\cal A}-\nabla\varphi$ leaves $\rho_c$ invariant, thereby recovering the usual notion of gauge-covariance. In this sense, the KvH construction is gauge-covariant. More specifically, we observe that the KvH Lagrangian in \eqref{VP-KvH} is \emph{manifestly covariant} in the sense that it satisfies the relation
\beq
L(e^{i\varphi/\hbar}\chi,e^{i\varphi/\hbar}\partial_t\chi;{\color{black}\boldsymbol{\cal A}-\nabla\varphi})=L(\chi,\partial_t\chi;\boldsymbol{\cal A})
\,,
\label{Covariance}
\eeq
where the semicolon indicates that the vector potential $\boldsymbol{\cal A}$ appears in the Lagrangian merely as  a parameter. In more generality, we recall that a \textit{covariant} Lagrangian allows for an additive total time derivative in the right-hand side of \eqref{Covariance}.  Such an additive time derivative is also allowed in the case of \textit{gauge-invariant} Lagrangians, while \emph{manifest gauge-invariance} simply means $L(e^{i\varphi/\hbar}\chi,e^{i\varphi/\hbar}\partial_t\chi)=L(\chi,\partial_t\chi)$. We notice that, while the KvN expression of the Liouville density is indeed manifestly gauge-invariant, the KvN Lagrangian in \eqref{VP-KvN1} is not even covariant. This is a point which we will come back to in Section \ref{sec:altvar}, where KvN theory will be endowed with a manifestly gauge-invariant Lagrangian.

Besides these technical details, we emphasize that this form of gauge transformations have been discussed in several prominent works in hybrid quantum-classical dynamics. In particular, here we refer to \cite{boucher,Ghose}. Therein, the authors point out that the gauge invariance under local $U(1)$ transformations is absolutely essential in reflecting the correct dynamics in the classical sector of hybrid systems. This is due to the fact that, as a general rule, phases are normally considered unobservable in classical mechanics. For example, as discussed in  \cite{Ghose}, Sudarshan's early approach to hybrid dynamics aimed at incorporating gauge invariance by enforcing  appropriate superselection rules. However, the role of this superselection rules has attracted a certain amount of criticism \cite{Barcelo,PeTe,Terno}. 

In this work superselection rules are replaced by local phase invariance, much in the same spirit as global phase invariance in standard quantum mechanics. In particular,  gauge invariance is achieved by starting with a   covariant formulation based on KvH theory, so that the full invariance can be enforced at the level of the underlying variational principle. The next section illustrates this process for the special case of purely classical dynamics.

\section{From KvH back to KvN\label{sec:BackToKvN}}

So far, we have observed that KvH theory overcomes the issues emerging in the standard variational formulation of KvN dynamics. However, one may insist that the KvN  prescription $\rho=|\chi|^2$ is  perfectly sensible  so that the question becomes to identify an alternative variational formulation overcoming the issues emerging from to the canonical structure for the KvN theory presented in Section \ref{sec:KvN}. {\color{black}While this alternative construction was presented in \cite{TrJo21}, here we will show how this can be obtained as a closure model emerging from KvH theory. This new result will pave the way to the construction of a hybrid quantum-classical model later on}.

We start our discussion by presenting a remark that motivates the steps taken later on. We realize that, if we could set $\partial_p S=0$ and $p=\partial_q S$ in \eqref{giggia}, the KvH expression of the classical density would recover the original KvN prescription $\rho_c=|\chi|^2$. Importantly, upon taking the gradient of the second in \eqref{KvH2Mad}, we notice that $\nabla S-\boldsymbol{\cal A}$ satisfies the evolution equation
\beq
\partial_t(\nabla S-\boldsymbol{\cal A})+\bX_H\cdot\nabla(\nabla S-\boldsymbol{\cal A})+\nabla \bX_H\cdot(\nabla S-\boldsymbol{\cal A})=0
\label{dS-A}
\eeq
so that the initial condition $\nabla S=\boldsymbol{\cal A}$ is preserved in time. However, this type of initial condition requires introducing topological singularities that are difficult to treat. Nevertheless, it is rather suggestive that replacing $\nabla S\to\boldsymbol{\cal A}$ takes the classical density $ \rho  _c$ back to to the KvN prescription $ \rho  _c=| \chi | ^2 $.  In what follows we will take full advantage of this observation, yet avoiding topological singularities.

\subsection{KvH variational principles and hydrodynamic paths\label{sec:EPMad}}
Instead of allowing for a singular phase, here we will pursue an alternative direction which consists in replacing $\nabla S\to\boldsymbol{\cal A}$ within a suitable form of the KvH variational principle. Indeed, in order to make this replacement, we first need to bring the variational principle \eqref{VP-KvH} into a form that is amenable to appropriate manipulations. For this purpose, we first express the Lagrangian $L_{\rm KvH}$ in terms of phase and amplitude, that is
\beq
L_{\rm KvH}=\int\!\Big(D\partial_t S-D\big(\{H,S\}+p\partial_pH- H\big)\Big)\,\de q\de p
\,.
\label{KvKMadLagr}
\eeq

Inspired by Madelung's hydrodynamic method in quantum mechanics, we will now introduce the characteristics of the vector field $\bX_H$. More specifically, borrowing the terminology from continuum dynamics, we introduce the {\it Lagrangian coordinate} ${\boldsymbol\eta}(q_0,p_0,t)$ such that $\dot{\boldsymbol\eta}(q_0,p_0,t)=\bX_H({\boldsymbol\eta}(q_0,p_0,t))$. 
 In this hydrodynamic picture, $\bX_H$ is regarded as the {\it Eulerian velocity} corresponding to its Lagrangian counterpart $\dot{\boldsymbol\eta}$. Then, as a consequence of the first equation in \eqref{KvH2Mad}, the density $D$ is transported along ${\boldsymbol\eta}$. Upon introducing the notation ${\boldsymbol{z}}=(q,p)$, the density transport can be written as the \emph{Lagrange-to-Euler map}
\beq
D({\boldsymbol{z}},t)=\int\!D_0({\boldsymbol{z}_0})\delta({\boldsymbol{z}}-{\boldsymbol\eta}({\boldsymbol{z}_0},t))\,\de^2 {{z}}_0=\frac{D_0({\boldsymbol{z}_0})}{{\cal J}_{\boldsymbol\eta}({\boldsymbol{z}_0},t)}\bigg|_{{\boldsymbol{z}_0}={\boldsymbol\eta}({\boldsymbol{z}},t)^{-1}}.
\label{LtoEmap}
\eeq
Notice that here we are retaining the Jacobian ${\cal J}_{\boldsymbol\eta}=\operatorname{\sf det}\nabla{\boldsymbol\eta}$ in the denominator for later convenience, although the flow of a Hamiltonian vector field identifies a canonical transformation with unit Jacobian.

Let us now replace \eqref{LtoEmap} in the Lagrangian $L_{\rm KvH}$.  In particular, let us apply the product rule $D\partial_t S=\partial_t(DS)-S\partial_tD$ to the first term of  \eqref{KvKMadLagr} and ignore the irrelevant total derivative $\partial_t(DS)$. Here, the time derivative $\partial_tD$ may be found from \eqref{LtoEmap} so that we have
\beq
\partial_tD+\operatorname{div}(\boldsymbol{\cal X}D)=0
\,,
\label{D-eq}
\eeq
with $\boldsymbol{\cal X}({\boldsymbol{z}})=\dot{\boldsymbol\eta}({\boldsymbol{z}_0},t)|_{{\boldsymbol{z}_0}={\boldsymbol\eta}^{-1}({\boldsymbol{z}},t)}$.
%
Notice that, while the original relation $\dot{\boldsymbol\eta}=\bX_H({\boldsymbol\eta})$ leads to $\bX_H({\boldsymbol{z}})=\dot{\boldsymbol\eta}({\boldsymbol{z}_0},t)|_{{\boldsymbol{z}_0}={\boldsymbol\eta}^{-1}({\boldsymbol{z}},t)}$,  the prescribed vector field $\bX_H$ has been replaced by the unknown vector field $\boldsymbol{\cal X}$. Indeed, in the present treatment the Lagrangian coordinate ${\boldsymbol\eta}$ is left as a dynamical variable and analogously for its velocity $\dot{\boldsymbol\eta}$. Then, the associated Eulerian vector field $\boldsymbol{\cal X}$ must be treated in the same way. 
With these steps in mind, we observe that the Lagrangian $L_{\rm KvH}$ in \eqref{KvKMadLagr} assumes the new form
\beq
L_{\rm EPKvH}(\boldsymbol{\cal X}, D, S) =\int\!D\Big(\nabla S\cdot\boldsymbol{\cal X}+(\boldsymbol{\cal A}-\nabla S)\cdot J\nabla H- H\Big)\,\de^2z
\,,
\label{EP-KvH}
\eeq
up to an ignorable total time derivative. For the sake of clarity, here we have indicated the functional dependence of  $L_{\rm EPKvH}$ on the dynamical variables.
We notice that the variational principle associated to this  Lagrangian is not of standard type in that, unlike $\delta S$, the variations $\delta D$ and $\delta\boldsymbol{\cal X}$ are not arbitrary. Instead, these variations must be found from their definitions in terms of $\boldsymbol\eta$. This procedure is a result of a reduction process from Lagrangian to Eulerian variables that is known as {\it Euler-Poincar\'e reduction} \cite{HoMaRa98}, which justifies the letters EP in the subscript `EPKvH' above. Once the KvH Lagrangian is taken into the Euler-Poincar\'e form \eqref{EP-KvH}, the equations of motion are obtained by simply adopting standard methods in Euler-Poincar\'e variational principles \cite{HoScSt09,MaRa98}. In this approach, equation \eqref{D-eq} appears as an auxiliary equation following from the definition \eqref{LtoEmap} and accompanying the variational principle $\delta\int_{t_1}^{t_2}\!L_{\rm EPKvH}\,\de t=0$.

One can verify that the variational relation
\beq
\delta D=-\operatorname{div}(D\boldsymbol{\cal Y})
\label{Dvar}
\eeq
follows from taking variations of the Lagrange-to-Euler map \eqref{LtoEmap}, upon defining the arbitrary displacement vector field
$
\boldsymbol{\cal Y}(\bz,t)=\delta{\boldsymbol\eta}({\boldsymbol{z}_0},t)|_{{\boldsymbol{z}_0}={\boldsymbol\eta}^{-1}({\boldsymbol{z}},t)}
$. By proceeding analogously, a more cumbersome vector calculus exercise leads to
\beq
\delta\boldsymbol{\cal X}=\partial_t\boldsymbol{\cal Y}+\boldsymbol{\cal X}\cdot\nabla\boldsymbol{\cal Y}-\boldsymbol{\cal Y}\cdot\nabla\boldsymbol{\cal X}
\,.
\label{Xvar}
\eeq
We verify that the variations $\delta\boldsymbol{\cal X}$ and $\delta D$ produce \eqref{dS-A}, which  is equivalent to the second equation in \eqref{KvH2Mad} up to an irrelevant number. In addition, arbitrary variations $\delta S$ lead to $\operatorname{div}(D\boldsymbol{\cal X})=\operatorname{div}(D \bX_H)$ so that  from \eqref{D-eq} one recovers the first equation in \eqref{KvH2Mad}. Thus, Hamilton's principle $\delta\int_{t_1}^{t_2}L_{\rm EPKvH}\,\de t=0$ with the Lagrangian \eqref{EP-KvH} does indeed recover  KvH dynamics in the form \eqref{KvH2Mad}.

\subsection{Alternative variational approach to KvN theory\label{sec:altvar}}
Following the observations preceding Section \ref{sec:EPMad}, let us now perform the replacement
\[
\nabla S\to\boldsymbol{\cal A}
\]
in the Euler-Poincar\'e Lagrangian \eqref{EP-KvH}.
We obtain
\beq
L_{\rm EPL} =\int\!D(\boldsymbol{\cal A}\cdot\boldsymbol{\cal X}- H)\,\de^2z
\,.
\label{EPLiouvLagr}
\eeq
Here, the subscript `EPL' anticipates the next result: the replacement $\nabla S\to\boldsymbol{\cal A}$ recovers the {\it E}uler-{\it P}oincar\'e variational formulation of the classical {\it L}iouville equation. Indeed, upon recalling that $\boldsymbol{\cal A}=(p,0)$ is constant in time, the variations \eqref{Dvar} and \eqref{Xvar} produce $\boldsymbol{\cal X}=\bX_H$ so that the auxiliary equation $\partial_t D+\operatorname{div}(D\boldsymbol{\cal X})=0$ becomes the classical Liouville equation. Notice that, upon writing $\boldsymbol{\cal X}$ in terms of its velocity and force components, that is $\boldsymbol{\cal X}(\bz,t)=(\mathpzc{u}(\bz,t),\mathpzc{f}(\bz,t))$, explicitly one has $\boldsymbol{\cal A}\cdot\boldsymbol{\cal X}=p \mathpzc{u}(q,p,t)$. However, it may be convenient to retain the full $\boldsymbol{\cal A}-$notation since the symplectic potential may acquire different suitable forms depending on the problem under consideration. For example, as noticed in \cite{BoGBTr19}, the symplectic potential $\boldsymbol{\cal A}(\bz)={J}\bz/2$ is particularly useful in the case of quadratic Hamiltonians.

At this point, as presented extensively in \cite{TrJo21}, one can perform the replacement $D=|\chi|^2$. Taking the square root of \eqref{LtoEmap} yields  the following Lagrange-to-Euler map for Koopman-von Neumann wavefunctions:
\beq
\chi({\boldsymbol{z}},t)=\frac{\chi_0({\boldsymbol{z}_0})}{\sqrt{{\cal J}_{\boldsymbol\eta}({\boldsymbol{z}_0},t)}}\bigg|_{{\boldsymbol{z}_0}={\boldsymbol\eta}^{-1}({\boldsymbol{z}},t)}
.
\label{LtoEmapchi}
\eeq
Consequently, an alternative KvN Lagrangian to the one in \eqref{VP-KvN1} may be written as
\beq
L_{\rm EPKvN} =\int\!|\chi|^2(\boldsymbol{\cal A}\cdot\boldsymbol{\cal X}- H)\,\de^2z
\,,
\label{EPKvNLagr}
\eeq
which is  manifestly gauge invariant under phase transformations $\chi\mapsto e^{i\varphi/\hbar}\chi$.
Here, the time derivative $\partial_t \chi$ and the variations $\delta\chi$ follow from \eqref{LtoEmapchi} and read
\beq\label{chivar}
\partial_t\chi=-\boldsymbol{\cal X}\cdot\nabla\chi-\frac12\chi\operatorname{div}\boldsymbol{\cal X}
\,,\qquad\qquad
\delta\chi=-\boldsymbol{\cal Y}\cdot\nabla\chi-\frac12(\operatorname{div}\boldsymbol{\cal Y})\chi
\,,
\eeq
respectively. 
The variations $\delta\boldsymbol{\cal X}$ are still given by \eqref{Xvar} and they produce the relation $\boldsymbol{\cal X}=\bX_H$, so that the first equation in \eqref{chivar} recovers the KvN equation in the form $\partial_t\chi=\{H,\chi\}$. In the general case, the first equation in \eqref{chivar} identifies the transport equation for a wavefunction, so that the well-known density transport in \eqref{D-eq}  is indeed recovered by setting $D=|\chi|^2$.
In this construction, no ambiguity occurs when the Koopman phase is set to zero and the total energy coincides with the KvN prescription $\int\! H|\chi|^2\,\de^2z$.  

{\color{black}Before concluding this section, we  present the noncanonical Hamiltonian structure of the KvN equation, as it arises from the variational principle \eqref{EPKvNLagr}. In general, a Hamiltonian structure comprises a Hamiltonian functional $h=h(\chi)$ and a Poisson bracket $\{\,,\,\}_\chi$, so that any functional $f(\chi)$ evolves according to $\de{f}/\de t=\{f,h\}_\chi$. A Poisson bracket  is a Lie bracket  satisfying the Leibniz product rule $\{f,gk\}_\chi=\{f,g\}_\chi k+\{f,k\}_\chi g$. The main difficulty in constructing a Poisson bracket consists in ensuring the Jacobi identity for cyclic permutations. There are lots of examples in the literature where candidate Poisson brackets fail to satisfy the Jacobi identity; see \cite{Se05} for a discussion in the context of hybrid quantum-classical dynamics. Thus, the safest way to construct a Poisson bracket is usually to derive it from another bracket which is already known to be Poisson. As discussed in \cite{GBTr21,TrJo21}, the Poisson bracket and Hamiltonian functional associated to the KvN equation read, respectively,
\beq
\{f,k\}_\chi=\int\!\frac1{|\chi|^2}\left(\chi\diamond\frac{\delta f}{\delta \chi}\right)\cdot{J}\left(\chi\diamond\frac{\delta k}{\delta \chi}\right)\de^2z
\,,\qquad\qquad
h(\chi)=\int\!H|\chi|^2\,\de^2z
\,,
\label{KvNHS2}
\eeq 
where we introduce 
\beq
\chi\diamond\frac{\delta f}{\delta \chi}:=\frac12
\operatorname{Re}\!\left(\frac{\delta h}{\delta \chi}^{\!*}\nabla\chi-\chi^*\nabla\frac{\delta f}{\delta \chi}\right)
\label{diamond1}
\eeq
for compactness of notation. As we will see, an analogue of this Poisson structure will reappear in the context of hybrid quantum-classical dynamics.
}

While this section has presented a prescription for obtaining an alternative variational structure for KvN theory that is immune of the ambiguities discussed in Section \ref{sec:KvN}, little was said about how this prescription can be justified within KvH theory. As already mentioned, the equality $\nabla S=\boldsymbol{\cal A}$ requires a singular phase which we want to avoid in the present work. Thus, the present wavefunction treatment is insufficient {\color{black}by itself} to obtain the KvN construction as a type of reduced model of KvH dynamics. An extension of this treatment that overcomes this difficulty is presented in Appendix \ref{sec:KoopMixtures}, {\color{black}which shows how KvN can be realized as an exact solution of the KvH equation for density matrices}.

\section{Hybrid quantum-classical dynamics\label{sec:hybrids}}

Having characterized the different Koopman formulations of classical mechanics, we move on to discussing the Koopman approach to mixed quantum-classical systems. In particular, here we will review a recent approach proposed by the authors and based on the KvH construction \cite{BoGBTr19,GBTr20}. This approach is heavily inspired by Sudarshan's early work \cite{Sudarshan}, which however was based on KvN theory. 
As we have seen in Section \ref{sec:KvH}, the KvH approach differs substantially from KvN by the fact that the former has a nontrivial phase dynamics. In the present context this phase dynamics is crucial in realizing the quantum-classical coupling.

While the approach discussed in this section succeeds in capturing essential properties such as quantum uncertainty, it is not clear whether the associated phase-space distribution of the classical subsystem remains positive in time. So far, this particular point was partly addressed in \cite{GBTr20}, where we identified an infinite family of hybrid systems for which the classical subsystem has a positive Liouville density at all times. However, a general statement is still lacking. Later on in this paper we will overcome this potential issue by presenting a closure model ensuring a positive classical density. At this stage, we  simply review the original approach which our closure model is based on.



\subsection{Hybrid wavefunctions\label{sec:hybWF}} 
As explained in \cite{BoGBTr19}, the interaction  of a quantum and a classical particle can be formulated by starting with the KvH equation for the two-particle wavefunction $\Upsilon(\bz_1,\bz_2)$ and then applying canonical quantization to quantize one of them. This approach was also partly followed in \cite{JaSu10}. The quantization process follows the usual rules: in particular, we make the replacements $q_2\to\hat{x}=x$ and $p_2\to\hat{p}=-i\hbar\partial/\partial x$. Then, upon setting $\partial\Upsilon/\partial p_2=0$ and dropping the subscripts, we obtain the \emph{quantum-classical wave equation}:
\begin{equation}\label{hybrid_KvH}
{\rm i}\hbar\partial_t\Upsilon=\{{\rm i}\hbar \widehat{H},\Upsilon\} + \big(\widehat{H} - p\partial_p{\widehat{H}}\big)\Upsilon
\,.
\end{equation}
Here, $\Upsilon(q,p,x)$ is a hybrid wavefunction, that is a square-integrable function of both the classical and quantum coordinates $\bz=(q,p)$ and $x$, respectively. Also, $\widehat{H}(q,p,\hat{x},\hat{p})$ is an operator-valued function on the classical phase-space. Following the discussion in Sections \ref{sec:KvN} and \ref{sec:KvH}, here we can write the equation above as
\beq
{\rm i}\hbar\partial_t\Upsilon=\widehat{\cal L}_{\widehat{H}}\Upsilon
\,,\qquad\text{ with }\qquad
\widehat{\cal L}_{\widehat{H}}:=\{ i\hbar \widehat{H},\, \}-p\partial_p\widehat{H}+\widehat{H}.
\label{QCWE}
\eeq
The \emph{quantum-classical Liouvillian operator} $\widehat{\cal L}_{\widehat{H}}$ is again Hermitian so that, similarly to quantum and classical mechanics, hybrid  dynamics is described as a unitary flow on the space of quantum-classical wavefunctions.

The hybrid wave equation \eqref{hybrid_KvH} possesses a profound geometric structure in symplectic geometry as discussed in \cite{GBTr20}. While here we will not enter the details, we will however emphasize that equation \eqref{hybrid_KvH} possesses a Dirac-Frenkel variational principle of the type \eqref{VP-KvN1} and \eqref{VP-KvH}. In particular, equation \eqref{hybrid_KvH} arises as the Euler-Lagrange equation associated to the Lagrangian
\beq
L_{\rm QC}(\Upsilon,\partial_t\Upsilon)=\operatorname{Re}\!\int\!(i\hbar\Upsilon^*\partial_t\Upsilon-\Upsilon^*\widehat{\cal L}_{\widehat{H}}\Upsilon)\,\de^2z\,\de x
\label{VP-Hyb1}
\,,
\eeq
where the last term identifies the total energy $h( \Upsilon )= \operatorname{Re} \int\Upsilon^*\widehat{\cal L}_{\widehat{H}}\Upsilon\,\de^2z\,\de x$.

\subsection{Quantum and classical densities\label{sec:hybden1}}
The Lagrangian \eqref{VP-Hyb1} is important in that its associated total energy naturally identifies a distribution-valued von Neumann operator playing the role of a hybrid quantum-classical density. Indeed, integration by parts yields 
\beq\label{toten}
\operatorname{Re}\!\int\!\Upsilon^*\widehat{\cal L}_{\widehat{H}}\Upsilon\,\de^2z\,\de x=\operatorname{\sf Tr}\!\int\!\widehat{H}\widehat{\cal D}\,\de^2z
\eeq 
with
\beq\label{hybden}
\widehat{\cal D}(\bz):=\Upsilon(\bz)\Upsilon^\dagger(\bz)+\partial_p(p\Upsilon(\bz)\Upsilon^\dagger(\bz))+i\hbar\{\Upsilon(\bz),\Upsilon^\dagger(\bz)\}
\,.
\eeq
Here, `$\dagger$' denotes the quantum adjoint so that, for example, the measure-valued operator  $\Upsilon(\bz)\Upsilon^\dagger(\bz)$ has matrix elements $\Upsilon(\bz,x)\Upsilon^*(\bz,x')$, while $\Upsilon^\dagger(\bz)\Upsilon(\bz)=\int |\Upsilon(\bz,x)|^2\,\de x$ is a phase-space density. The right-hand side of the relation \eqref{toten} confers the distribution-valued operator $\widehat{\cal D}(\bz)$ the role of a hybrid density for the calculation of expectation values. Given a hybrid observable represented by an operator-valued function $\widehat{A}(\bz)$, we prescribe the following definition of its expectation value:
$\langle \widehat{A}\rangle=\operatorname{\sf Tr}\int\widehat{\cal D}(\bz)\widehat{A}(\bz)\,\de^2z$. The hybrid Ehrenfest theorem for expectation value dynamics was presented in \cite{BoGBTr19}, where we showed that the total quantum-classical momentum $\langle p+\hat{p}\rangle$ remains conserved in time for translation-invariant Hamiltonians. Recent work \cite{Andre} has investigated  this result in the context of the Galilean covariance of the quantum-classical wave equation \eqref{QCWE}.

We notice that the hybrid density operator \eqref{hybden} is sign-indefinite.
Despite this potential issue, we observe that the quantum density matrix
\beq
\hat{\rho}=\int\!\widehat{\cal D}(\bz)\,\de^2z = \int \Upsilon(\bz)\Upsilon^\dagger(\bz)\,\de^2z
\label{qden}
\eeq
is actually positive-definite by construction and this property ensures that the Heisenberg uncertainty principle is satisfied. This is a crucial ingredient in the study of quantum-classical coupling. For example, the AG equation \eqref{AGeq} fails to satisfy this property. To see how decoherence is captured in the present hybrid context, we write the dynamics of the quantum density matrix as \cite{BoGBTr19}
\beq\label{quantevol}
i\hbar\frac{\partial \hat\rho}{\partial t}=\int[\widehat{H},\widehat{\cal D}]\,\de^2z
\,.
\eeq
In general, this equation does not allow for an initial pure quantum state $\hat{\rho}_0=\psi_0\psi_0^\dagger$ to remain pure over time. Indeed, upon defining the quantum purity as $\operatorname{\sf Tr}(\hat\rho^2)$, we see that $\de\operatorname{\sf Tr}(\hat\rho^2)/\de t\neq0$ thereby leading to quantum decoherence. As we will see below,  purity non-preservation also affects the classical evolution. 

Let us look closely at the expression of the classical Liouville distribution, that is
\beq
\rho_c(\bz)=\operatorname{\sf Tr}\widehat{\cal D}(\bz)=\int\!\Big(|\Upsilon(\bz)|^2+\partial_p(p|\Upsilon(\bz)|^2)+\hbar\operatorname{Im}\{\Upsilon^*(\bz),\Upsilon(\bz)\}\Big)\,\de x
\,.
\label{cden}
\eeq
Importantly, this quantity  is generally sign-indefinite. At present, we do not know whether an initially positive $\rho_c$ stays always positive in time or may develop negative values. The evolution on the classical distribution $\rho_c$ is governed by the equation
\beq
\frac{\partial \rho_c}{\partial t}=\operatorname{\sf Tr}\{\widehat{H},\widehat{\cal D}\}
\,.
\label{classevol}
\eeq
So far, we could only prove that this equation does not change the initial sign of $\rho_c$  whenever the hybrid Hamiltonian $\widehat{H}$ depends  only on a set of mutually commuting operators; see \cite{GBTr20} for further details. This property was confirmed in \cite{BoGBTr19} for the case of a pure-dephasing problem with quadratic coupling. In this particular case, we observed that both the quantum density \eqref{qden} and the classical density \eqref{cden} remain positive in time while the hybrid density \eqref{hybden} develops negative eigenvalues.

As the question whether \eqref{cden} remains always positive is currently open, one may use analogies with Wigner functions to justify the possible emergence of negative values in the classical distribution in the presence of quantum-classical interaction \cite{BoGBTr19}. Alternatively, one may want to enforce a positive Liouville density in some way. We will pursue this second direction later on in this paper. 

Before closing this section, we observe that equation \eqref{classevol} does not allow for Klimontovich particle solutions. Specifically, a $\delta-$like initial condition $\rho_{c\,0}(\bz)=\delta(\bz-\bzeta_0)$ gets spread across phase-space as time goes by. While this may sound somewhat reminiscent of diffusion processes, we emphasize that no diffusion mechanism takes place in this case; at least, not in any standard sense. Since Klimontovich states are classical pure states \cite{ChMa76,Sh79,Tronci2018}, we conclude that this classical form of purity non-preservation is analogous to the decoherence appearing in the quantum sector. This point raises the foundational question whether this phenomenon in the classical sector is related to the uncontrollable disturbance invoked by Bohr in his original interpretation of quantum measurements. 

Also, we notice that while the loss of classical pure states leads to a form of statistical uncertainty, the latter is very different from Heisenberg's relation $\Delta \langle \hat{x}\rangle\Delta\langle \hat{p}\rangle\geq \hbar/2$  which establishes a well-characterized lower bound. Thus, the form of classical decoherence experienced by the classical system cannot be considered as a loss of its `classicality'; instead, it simply means that  this system must be described by a statistical phase-space distribution in  the standard Liouville picture.


\section{Exact factorization of hybrid wavefunctions\label{sec:EF}}

In this section, we formulate a closure scheme of the  quantum-classical wave equation \eqref{QCWE} so that \textit{both the quantum and classical densities  are positive at all times}. While the quantum density operator \eqref{qden} is already positive-definite by construction, we will ensure a positive classical density by exploiting a gauge-invariance principle, as outlined in the Introduction. In particular, we will modify the original theory from Section \ref{sec:hybrids} to enforce a gauge symmetry with respect to local phase factors for which the phase is a function of the classical phase space. For the purely classical case, a similar procedure was followed in Sections \ref{sec:EPMad} and \ref{sec:altvar}. However, before this procedure can be applied in the hybrid context, we face the difficulty that the classical degrees of freedom cannot be  isolated in a simple way since the current form of the theory involves a hybrid wavefunction $\Upsilon(\bz,x)$ in which quantum and classical coordinates, respectively $x$ and $\bz=(q,p)$, are treated on an equal footing.

In order to circumvent this difficulty, we resort to a method from chemical physics \cite{AbediEtAl2012}. Known under the name \emph{exact factorization}, in our context this method simply consists in rewriting the hybrid wavefunction as follows:
\beq\label{EFDef}
\Upsilon(\bz,x,t)=\chi( \bz,t)\psi(x,t;\bz)\,,
\qquad\text{with}\qquad
\int|\psi(x,t;\bz)|^2\,\de x=1
\,,
\eeq
where the semicolon indicates that the classical coordinates appear in the expression of the quantum wavefunction $\psi$ merely as parameters. In other words, one has a  Koopman  wavefunction $\chi(\bz,t)$ and a  Schr\"odinger  wavefunction $\psi(x,t;\bz)$, where the latter is parameterized by the classical coordinates $\bz=(q,p)$. While so far the quantum state space was identified with the  infinite-dimensional space $L^2(\Bbb{R})$ of square-integrable Schr\"odinger wavefunctions, in what follows we shall consider an arbitrary quantum Hilbert space $\mathscr{H}_{\scriptscriptstyle Q}$ with inner product $\langle\cdot|\cdot\rangle$, induced norm $\|\!\cdot\!\|$,  and real-valued pairing 
\beq
\langle\cdot,\cdot\rangle=\operatorname{Re}\langle\cdot|\cdot\rangle
\,.
\label{pairing} 
\eeq
For later purpose, it is convenient to introduce the shorthand notation 
\beq
\langle\widehat{A}\rangle=\langle\psi,\widehat{A}\psi\rangle
\label{electexp}
\eeq
for any Hermitian operator-valued function on phase-space  $\widehat{A}=\widehat{A}(\bz)$. We hope that no confusion arises with the expectation value notation introduced in Section \ref{sec:hybden1}.
Notice that both  wavefunctions $\chi$ and $\psi$ in the factorization \eqref{EFDef} are  defined up to an arbitrary phase factor $e^{i\varphi(t,\bz)}$. 

The name \emph{exact factorization} arises from the fact that the relation \eqref{EFDef}  generally identifies an  exact solution of \eqref{hybrid_KvH}, as long as $\chi$ is nowhere vanishing. Over the years, factorizations of the type \eqref{EFDef}  appeared  in the context of standard quantum mechanics. For example, this is the common approach to the hydrodynamic formulation of the Pauli equation \cite{BBirula}. However, it was only in \cite{AbediEtAl2012} that this type of wavefunction factorization was recognized to have more general validity. The geometric underpinning of \eqref{EFDef} was studied in \cite{FoHoTr19,FoTr,HoRaTr21} and the results therein provide the basis for the present work.

Importantly, we observe that, upon writing $\chi=\sqrt{D}e^{iS/\hbar}$, the exact factorization \eqref{EFDef} transforms the original expression \eqref{cden} into the form
\beq\label{densities2}
\rho_c=D+\operatorname{div}\!\big(DJ(\nabla S+\boldsymbol{\cal A}_B-\boldsymbol{\cal A})\big),
\eeq
{\color{black}where we have introduced the \emph{Berry connection}
\beq\label{berryconn}
\boldsymbol{\cal A}_B:=\langle\psi,-i\hbar\nabla\psi\rangle\,.
\eeq}
Thus, similarly to the arguments in Section \ref{sec:BackToKvN}, we realize that setting 
\beq\label{closrel}
\nabla S+\boldsymbol{\cal A}_B=\boldsymbol{\cal A}
\eeq
would lead to a positive definite classical density $\rho_c=|\chi|^2$.  Motivated by this observation, we will perform various steps in order to apply the method in Sections \ref{sec:EPMad} and \ref{sec:altvar} within the present hybrid setting. As we will see, the relation \eqref{closrel} leads to a gauge-invariant quantum-classical theory for which classical phases merely represent a gauge freedom and are thus unobservable.

The discussion in the following sections provides an alternative treatment to that in \cite{GBTr21}, where the same closure model was obtained in terms of density matrices by exploiting the mathematical methods from geometric mechanics \cite{HoScSt09,MaRa98}. Instead, the present treatment follows a more direct approach mostly based on wavefunctions.

\subsection{Madelung transform and variational approach\label{sec:HydFrame}}

As we aim to develop a closure model by using the tools illustrated in Section \ref{sec:BackToKvN}, it is important to combine the variational approach to the Madelung transform with the Euler-Poincar\'e construction, as presented in Section \ref{sec:EPMad}. Here we restrict to focus only on the classical degrees of freedom and thus we choose to apply a partial Madelung transform only to the classical part of the wavefunction in \eqref{EFDef}. Thus, upon writing $\chi=\sqrt{D}e^{iS/\hbar}$, we replace the ansatz $\Upsilon(\bz,x)=\sqrt{D(\bz)}e^{iS(\bz)/\hbar}\psi(x;\bz)$ into the hybrid Lagrangian \eqref{VP-Hyb1}, thereby obtaining
\beq
L_{\rm EF}=\int \!D\big(\partial_t S-\langle\psi,i\hbar\partial_t\psi\rangle
\big)\,\de^2z+h(D,S,\psi)
\,,
\label{Tiziana}
\eeq
where the subscript `EF' stands for Exact Factorization and the Hamiltonian $h(D,S,\psi)$ arising from \eqref{VP-Hyb1} is written as
\begin{align}\nonumber
h(D,S,\psi)=& \int \!D\Big(\big\langle\bX_{\widehat{H}}\big\rangle\cdot(\nabla S+\boldsymbol{\cal A}_B)+\big\langle \widehat{ \mathcal{L} }_{\widehat{H}}-\boldsymbol{\cal A}_B\cdot\bX_{\widehat{H}}\big\rangle\Big)\,\de^2z
\\
=&
\int \!D\Big(\big\langle\bX_{\widehat{H}}\big\rangle\cdot(\nabla S+\boldsymbol{\cal A}_B-\boldsymbol{\cal A})+\big\langle\psi,{\widehat{H}}\psi-i\hbar\widetilde{\bX}_{\widehat{H}}\cdot\nabla\psi\big\rangle\Big)\,\de^2z
\,.
\label{hamilt1}
\end{align}
 Note  that \eqref{Tiziana} is the quantum-classical extension of the Lagrangian \eqref{KvKMadLagr} for the classical case.
Here, we recall \eqref{electexp} and we have denoted
\beq
\bX_{\widehat{H}}=(\partial_p\widehat{H},-\partial_q\widehat{H})\,,
\qquad
\qquad
\widetilde{\bX}_{\widehat{H}}:={\bX}_{\widehat{H}}-\langle{\bX}_{\widehat{H}}\rangle
\,,
\label{alexandra}
\eeq
while $\widehat{ \mathcal{L} }_{\widehat{H}}$ was defined in \eqref{QCWE}. 
Also, all differential operators act on the classical phase-space and we  recall the symplectic potential $\boldsymbol{\cal A}=(p,0)$ from Section \ref{sec:BackToKvN}. More importantly, in \eqref{hamilt1} we have added and subtracted the {\color{black}{Berry connection} $\boldsymbol{\cal A}_B$}. In addition, we notice that the condition $\|\psi\|^2=1$ is not enforced here as a constraint. As we will see, this condition is preserved in time by the final equations of motion. 

Arbitrary variations $\delta S$ and $\delta D$ in the action functional associated to \eqref{Tiziana} yield
\begin{align}
&\partial_t D+\operatorname{div}(D\langle \bX_{\widehat{H}}\rangle)=0,
\label{mario2}
\\
&\partial_t S+\langle \bX_{\widehat{H}}\rangle\cdot\nabla S=\big\langle\psi,i\hbar\partial_t\psi\big\rangle-\big\langle\psi,\widehat{ \mathcal{L} }_{\widehat{H}}\psi\big\rangle
\label{S_equ}
\end{align}
and we will next focus on the equation for the quantum wavefunction $\psi$.

\subsection{Quantum wavefunction and classical flow paths\label{sec:QWFCFP}}
In order to write the quantum evolution equation, it is convenient to introduce the functional
\begin{align}
f(D, \psi ) :=&\int\!D\big\langle\psi, (\widehat{ \mathcal{L} }_{\widehat{H}}-\boldsymbol{\cal A}_B\cdot\bX_{\widehat{H}})\psi\big\rangle\,\de^2z
\,.
\label{ffunct}
\end{align}
Then, arbitrary variations $\delta\psi$ in $\int_{t_1}^{t_2}L_{\rm EF} \,{\rm d} t$ yield \cite{GBTr21}
\beq
\label{psieq2}
i\hbar  \big(\partial_t+\langle \bX_{\widehat{H}}\rangle\cdot \nabla \big)\psi=
(\nabla S+\boldsymbol{\cal A}_B)\cdot  \bX_{\widehat{H}}\psi+\frac1{2D}\frac{\delta f}{\delta\psi}
\,,
\eeq
where we compute
\[
\frac{\delta f}{\delta\psi}=2D\widehat{H}\psi-2D(\boldsymbol{\cal A}+\boldsymbol{\cal A}_B)\cdot\bX_{\widehat{H}}\psi- 2i\hbar D\widetilde{\bX}_{\widehat{H}}\cdot\nabla\psi- i\hbar \psi\operatorname{div}(D\widetilde{\bX}_{\widehat{H}})
\,.
\]
While the current explicit form of equation \eqref{psieq2} is not particularly insightful, we will be able to make relevant considerations by recognizing that the functional \eqref{ffunct} can be entirely expressed in terms of the local density matrix $\hat\rho(\bz)=\psi(\bz)\psi(\bz)^\dagger$ as follows:
\begin{align}
f=&\,\frac12\int \!D\Big(\big\langle\bX_{\widehat{H}},i\hbar[\hat\rho,\nabla\hat\rho]\big\rangle-2\big\langle \boldsymbol{\cal A}\cdot\bX_{\widehat{H}}-\widehat{H},\hat\rho\big\rangle\Big)\,\de^2z
\label{hannah}
\,.
\end{align}
Here, we use the real-valued pairing $\color{black}\langle\widehat{A},\widehat{B}\rangle=\operatorname{Re}\langle\widehat{A}|\widehat{B}\rangle=\operatorname{Re}\operatorname{\sf Tr}(\widehat{A}^\dagger\widehat{B})$.
Notice the appearance of the non-Abelian gauge connection $[\hat\rho,\nabla\hat\rho]$ already emerged in Mead's work on molecular geometric phases \cite{Me92}. More importantly, the chain-rule relation $\delta f/\delta\psi=2(\delta f/\delta\hat\rho)\psi$ allows us to rewrite  equation \eqref{psieq2} as
\beq
\label{psieq3}
i\hbar  \big(\partial_t+\langle \bX_{\widehat{H}}\rangle\cdot \nabla \big)\psi= \bigg((\nabla S+\boldsymbol{\cal A}_B)\cdot  \bX_{\widehat{H}}+\frac1D\frac{\delta f}{\delta\hat\rho}\bigg)\psi
\,,
\eeq
where we compute
\begin{align*}
\frac{\delta f}{\delta\hat\rho}
=
D\widehat{H}-D\boldsymbol{\cal A}\cdot\bX_{\widehat{H}}+
\frac{i\hbar}2\left(D\{\hat\rho,\widehat{H}\}+D\{\widehat{H},\hat\rho\}+\big[\hat\rho,\{D,{\widehat{H}}\}\big]\right).
\end{align*}
At this point, the form of equation \eqref{psieq3} reveals the structure underlying the quantum evolution. Indeed, we realize that, since $\delta f/\delta\hat\rho$ acts on $\psi(\bz,t)$ as a Hermitian operator and so does $(\nabla S+\boldsymbol{\cal A}_B)\cdot  \bX_{\widehat{H}}$, the parenthesis on the right-hand side of \eqref{psieq3} generates a unitary propagator parameterized by the phase-space coordinates. This implies $(\partial_t+\langle \bX_{\widehat{H}}\rangle\cdot \nabla )\|\psi\|^2=0$, so that  the partial normalization condition $\|\psi(\bz,t)\|^2=1$ is preserved in time. More importantly, we notice the appearance of the \emph{material derivative} $\partial_t+\langle \bX_{\widehat{H}}\rangle\cdot \nabla$ already present in \eqref{S_equ}. This indicates that the unitary quantum evolution occurs in a phase-space frame moving with the Lagrangian flow path $\boldsymbol\eta(\bz_0,t)$ generated by the vector field $\langle \bX_{\widehat{H}}\rangle$, which transports the density $D$ as in \eqref{mario2}. This important remark was recently exploited in \cite{GBTr21}. In the fully quantum case, a similar approach was previously followed in \cite{FoHoTr19}.

Having uncovered the hydrodynamic Lie-transport features underlying the exact factorization system, we are motivated to cast the Lagrangian \eqref{Tiziana} in Euler-Poincar\'e form, which is a more natural setting for hydrodynamic continuum theories. Following the arguments in Section \ref{sec:EPMad}, we consider the Lagrange-to-Euler map \eqref{LtoEmap} and  write the Lagrangian \eqref{Tiziana} in Euler-Poincar\'e form as
\beq
L_{\rm EPEF}=\int \!D\big(\nabla S\cdot\boldsymbol{\cal X}+\langle\psi,i\hbar\partial_t\psi\rangle\big)\,\de^2z-h(D,S,\psi)
\,,
\label{Tiziana2}
\eeq
where $h(D,S,\psi)$ is given in \eqref{hamilt1}. Note that \eqref{Tiziana2} is the quantum-classical extension of the Lagrangian \eqref{EP-KvH} for KvH classical mechanics.
We recall that the vector field $\boldsymbol{\cal X}$ is defined in terms of the Lagrangian phase-space path $\boldsymbol{\eta}(\bz_0,t)$ as $\dot{\boldsymbol{\eta}}(\bz_0,t)=\boldsymbol{\cal X}(\boldsymbol{\eta}(\bz_0,t),t)$, so that one has the constrained variations \eqref{Dvar}-\eqref{Xvar} along with arbitrary variations $\delta S$. One can verify that, up to an irrelevant time-dependent number, the Lagrangian \eqref{Tiziana2} leads to  equation \eqref{S_equ}, while the auxiliary equation $\partial_t D+\operatorname{div}(D\boldsymbol{\cal X})=0$ recovers \eqref{mario2}. In addition, the quantum evolution equation \eqref{psieq2} follows from taking arbitrary variations $\delta\psi$.

\color{black}
So far the exact factorization \eqref{EFDef} has not provided much insight into the nature of quantum-classical coupling, which still looks very intricate. However, as anticipated, the variational setting made available by the Lagrangian \eqref{Tiziana2}  allows us to apply the method in Section \ref{sec:BackToKvN} thereby leading to a closure model ensuring that the classical density \eqref{densities2} is positive definite at all times. This is done in the next section.

\section{A nonlinear hybrid wave equation\label{sec:NHWE}}

Let us now extend the method from Section \ref{sec:BackToKvN} to the hybrid setting and use that  as a tool for ensuring a  positive-definite classical density. 

\subsection{The quantum-classical Lagrangian\label{sec:HybLagrFun}}
In first place, 
we recall that the exact factorization \eqref{EFDef} transforms the original expression \eqref{cden} into \eqref{densities2}. Thus, here we want to be able to use the relation  \eqref{closrel} by following the method in Section \ref{sec:BackToKvN}. In this way, the classical density becomes $\rho_c=|\chi|^2$, which is positive-definite.
Notice that, unlike the classical case treated in Section \ref{sec:BackToKvN}, the relation \eqref{closrel} may not be preserved by the hybrid system \eqref{mario2}, \eqref{S_equ}, and \eqref{psieq2}. As we show in Appendix \ref{sec:vNOp}, the wavefunction treatment  can be extended to von Neuman operators, in which case relation \eqref{closrel} can be realized as a dynamical closure ansatz. In the present discussion, however, we will simply follow the procedure from Sections \ref{sec:EPMad} and \ref{sec:altvar}, which is entirely based on wavefunctions. In particular, we will make use of the relation \eqref{closrel} by replacing $\nabla S\to\boldsymbol{\cal A}-\boldsymbol{\cal A}_B
$ in the Lagrangian \eqref{Tiziana2}. Then, the latter  becomes  gauge-independent   and acquires the form
\beq
\ell(\boldsymbol{\cal X},D,\psi,\partial_t\psi)=\int \!D\big((\boldsymbol{\cal A}-\boldsymbol{\cal A}_B)\cdot\boldsymbol{\cal X}+\big\langle\psi,i\hbar\partial_t\psi-{\widehat{H}}\psi+i\hbar\widetilde{\bX}_{\widehat{H}}\cdot\nabla\psi\big\rangle\big)\,\de^2z
\,,
\label{Tiziana3}
\eeq
where the last two terms are found by replacing \eqref{closrel} in 
\eqref{hamilt1}. At this point, the relation \eqref{densities2} yields $\rho_c=D=|\chi|^2$, thereby recovering a positive classical density at all times. Indeed, we observe that the initial sign of $D$ is preserved by its dynamics, which is given by transport equation $\partial_t D+\operatorname{div}(D\boldsymbol{\cal X})=0$.

Upon using $D=|\chi|^2$, one may restore the hybrid quantum-classical wavefunction $\Upsilon=\chi\psi$. For this purpose, we notice that
$
\langle\Upsilon,-i\hbar\nabla\Upsilon\rangle
=
D\nabla S+D\langle\psi, - i\hbar\nabla\psi\rangle
$, 
where we recognize the appearance of the Berry connection \eqref{berryconn} in the right hand-side. Since the addition of the pure differential $\nabla S$ is irrelevant, we are led to writing
\beq\label{gigi}
\boldsymbol{\cal A}_B=\frac{\langle\Upsilon,-i\hbar\nabla\Upsilon\rangle}{\|\Upsilon\|^2}
\,.
\eeq
Here,  the angle brackets denote the quantum pairing, that is $\langle\Upsilon_1,\Upsilon_2\rangle=\operatorname{Re}(\Upsilon_1^\dagger(\bz)\Upsilon_2(\bz))$, so that the notation \eqref{electexp} leads to writing $\langle\widehat{A}\rangle= \langle \Upsilon , \widehat{A}  \Upsilon \rangle /\| \Upsilon \| ^2 $.
With this notation, the Lagrangian  \eqref{Tiziana3} becomes
\begin{align}
\ell_{\scriptscriptstyle QC}(\boldsymbol{\cal X},\Upsilon,\partial_t\Upsilon)=&\int\! \big\langle\Upsilon,i\hbar \partial_t\Upsilon+(\boldsymbol{\cal A}-\boldsymbol{\cal A}_B)\cdot\boldsymbol{\cal X}\Upsilon
\big\rangle\,\de^2z-h_{\scriptscriptstyle QC}(\Upsilon)
\,,
\label{untangledLagr}
\end{align}
where the quantum-classical Hamiltonian functional  now reads
\begin{align}
h_{\scriptscriptstyle QC}(\Upsilon)=& \int\langle\Upsilon,\widehat{\cal L}_{\widehat{H}}\Upsilon+(\boldsymbol{\cal A}-\boldsymbol{\cal A}_B)\cdot \bX_{\widehat{H}}\Upsilon
\rangle\,\de^2z
\nonumber
\\
=&\int\!\big\langle\Upsilon,{\widehat{H}}\Upsilon-i\hbar\widetilde{\bX}_{\widehat{H}}\cdot\nabla\Upsilon\big\rangle\,\de^2z\,.
\label{QCHam}
\end{align}
Here, the variation $\delta\Upsilon$ is arbitrary and $\delta \boldsymbol{\cal X}$ is given by \eqref{Xvar}.
Notice that, while the original hybrid Hamiltonian $h_{\scriptscriptstyle QC}( \Upsilon )=\int\langle\Upsilon,\widehat{\cal L}_{\widehat{H}}\Upsilon\rangle\,\de^2z\,\de x$ appearing in \eqref{VP-Hyb1} is manifestly covariant with respect to gauge transformations $\Upsilon(\bz)\mapsto e^{i\varphi(\bz)/\hbar}\Upsilon(\bz)$, the new reduced Hamiltonian \eqref{QCHam} is manifestly gauge invariant.  This  gauge invariance under local phase factors is a long-sought property in the theory of quantum-classical wavefunctions and its absence in previous hybrid models has been referred to as a ``severe problem'' \cite{boucher}. The fact that the present theory is immune from these issues, thereby ensuring  positive quantum and classical densities, represents an important step forward in the theory of quantum-classical coupling.

In addition, we observe that the total energy identified by the Hamiltonian functional \eqref{QCHam} clearly differs from the usual expression of the expectation value of the Hamiltonian operator, that is the first term $\int \langle\Upsilon,{\widehat{H}}\Upsilon\rangle\,\de^2z$. A possible interpretation of this fact is that  the energy balance of quantum-classical interactions involves an extra work -- here given by the second term in \eqref{QCHam} -- that is produced by correlation effects. This seems to be the possibility suggested by the authors of \cite{Carroll}. Alternatively, we may simply use integration by parts to isolate a  form of the hybrid quantum--classical density that is alternative to that presented in \eqref{hybden}. This second approach will be discussed in Section \ref{sec:HybDenOp}. Instead, the next section proceeds by writing the equation for $\Upsilon$.

\color{black}
\subsection{Hybrid wave equation\label{sec:NHWEeq}} 
At this point, we are ready to take variations and write the hybrid wave equation for $\Upsilon$. Indeed, taking arbitrary variations  $\delta\Upsilon$ in \eqref{untangledLagr} yields a nonlinear wave equation of the form 
\beq
i\hbar\frac{\partial \Upsilon}{\partial t}+i\hbar\boldsymbol{\cal X}\cdot\nabla\Upsilon+\frac{i\hbar}2\Upsilon\operatorname{div}\boldsymbol{\cal X}=\frac12\frac{\delta h_{\scriptscriptstyle QC}}{\delta\Upsilon}-\boldsymbol{\cal X}\cdot\boldsymbol{\cal A}\Upsilon
,
\label{prelimNHWE}
\eeq
where  we compute
\begin{align}
\frac12\frac{\delta h_{\scriptscriptstyle QC}}{\delta \Upsilon}
=&\, 
\big({\widehat{H}}
-\boldsymbol{\cal A}_B\cdot\widetilde{\bX}_{\widehat{H}}\big)\Upsilon-i\hbar\widetilde{\bX}_{\widehat{H}}\cdot\nabla\Upsilon-\frac12
i\hbar(\operatorname{div}\widetilde{\bX}_{\widehat{H}})\Upsilon
\,.
\label{funcder}
\end{align}
Here, we have used  \eqref{QCHam} and noticed that $ \delta \langle{\bX}_{\widehat{H}}\rangle/\delta\Upsilon  =  2\| \Upsilon \| ^{-2} { \widetilde{\bX}_{\widehat{H}}\Upsilon } $.
Then, equation \eqref{prelimNHWE} becomes
\beq
i\hbar\frac{\partial \Upsilon}{\partial t}+i\hbar(\boldsymbol{\cal X}+\widetilde{\bX}_{\widehat{H}})\cdot\nabla\Upsilon+\frac{i\hbar}2\operatorname{div}(\boldsymbol{\cal X}+\widetilde{\bX}_{\widehat{H}})\Upsilon=\big({\widehat{H}}
-\boldsymbol{\cal A}_B\cdot\widetilde{\bX}_{\widehat{H}}\big)\Upsilon
-\boldsymbol{\cal X}\cdot\boldsymbol{\cal A}\Upsilon
\,.
\label{NHWE}
\eeq
As a consequence, we observe that equation \eqref{prelimNHWE} leads to the following continuity equation for the classical Liouville density $\rho_c=\|\Upsilon\|^2$:
\beq
\partial_t\|\Upsilon\|^2+\operatorname{div}(\|\Upsilon\|^2\boldsymbol{\cal X})=0
\,.
\label{HybDenCont}
\eeq
This shows that the vector field $\boldsymbol{\cal X}$ indeed identifies the Lagrangian trajectories of the classical flow, which advances the classical density according to  \eqref{LtoEmap}, that is $\rho_c({\boldsymbol{z}},t)=\int\!\rho_{c0}({\boldsymbol{z}_0})\delta({\boldsymbol{z}}-{\boldsymbol\eta}({\boldsymbol{z}_0},t))\,\de^2 {{z}}_0$.

At this stage, however, the vector field $\boldsymbol{\cal X}$ still needs to be determined. In order to find its expression, we first see that taking variations $\delta \boldsymbol{\cal X}$ in \eqref{untangledLagr} according to \eqref{Xvar} and making use of \eqref{HybDenCont} yields
\begin{align}
\nonumber
\boldsymbol{\cal X}\cdot\nabla\boldsymbol{\cal A}
+\nabla\boldsymbol{\cal X}\cdot\boldsymbol{\cal A}
=&\ 
\partial_t\boldsymbol{\cal A}_B
+
\boldsymbol{\cal X}\cdot\nabla\boldsymbol{\cal A}_B
+\nabla\boldsymbol{\cal X}\cdot\boldsymbol{\cal A}_B
\\
=&\ 
\frac1{2\|\Upsilon\|^2}\bigg(\bigg\langle\frac{\delta h_{\scriptscriptstyle QC}}{\delta \Upsilon},\nabla\Upsilon\bigg\rangle-\bigg\langle\Upsilon,\nabla\frac{\delta h_{\scriptscriptstyle QC}}{\delta \Upsilon}\bigg\rangle\bigg)+\nabla(\boldsymbol{\cal X}\cdot\boldsymbol{\cal A})
\label{patrizia}
\,.
\end{align}
The second equality is easily verified. Indeed, upon introducing  $\gamma=\Upsilon/\|\Upsilon\|$, we use $\boldsymbol{\cal A}_B=\langle\gamma,-i\hbar\nabla\gamma\rangle$
to notice that \eqref{prelimNHWE} leads to $i\hbar\|\Upsilon\|({\partial_t}+\boldsymbol{\cal X}\cdot\nabla)\gamma=(1/2){\delta h_{\scriptscriptstyle QC}}/{\delta\Upsilon}-\boldsymbol{\cal X}\cdot\boldsymbol{\cal A}\Upsilon$. The latter can then be used to expand the terms after the first equality in \eqref{patrizia}.
Therefore, upon using the defining relation $(\nabla\boldsymbol{\cal A})^T-\nabla\boldsymbol{\cal A}=J$, we obtain 
\begin{align}
\boldsymbol{\cal X}=&\ \frac1{2\|\Upsilon\|^2}
\,J\bigg(\bigg\langle\Upsilon,\nabla\frac{\delta h_{\scriptscriptstyle QC}}{\delta \Upsilon}\bigg\rangle-\bigg\langle\frac{\delta h_{\scriptscriptstyle QC}}{\delta \Upsilon},\nabla\Upsilon\bigg\rangle\bigg)
\nonumber
\\\nonumber
=&\  \frac1{\|\Upsilon\|^2}J\bigg[\|\Upsilon\|^2\langle\nabla(\widehat{H}-\widetilde{\bX}_{\widehat{H}}\cdot\boldsymbol{\cal A}_B)\rangle+\boldsymbol{\cal A}_B\langle\Upsilon,(\operatorname{div}\widetilde{\bX}_{\widehat{H}})\Upsilon\rangle
\\
&\,\hspace{1.5cm} 
+
2\big\langle\nabla\Upsilon, \widetilde{\bX}_{\widehat{H}}\cdot(i\hbar\nabla)\Upsilon\big\rangle
-\nabla\big\langle\Upsilon,\widetilde{\bX}_{\widehat{H}}\cdot(i\hbar\nabla)\Upsilon\big\rangle
\bigg]
\label{vectfield}
\,,
\end{align}
where the second equality is proved in  Appendix \ref{Append1} using \eqref{funcder}. {\color{black}Alternatively, a further vector calculus exercise shows that
\begin{align}\nonumber
\boldsymbol{\cal X}
=&\,\langle \bX_{\widehat{H}}\rangle+\frac1{2}\Big(\big\langle (J\boldsymbol{\widehat{\Gamma}})\cdot\nabla,\bX_{{\widehat{H}}}\big\rangle
-\big\langle \{  \ln D,\widehat{H}\}+\bX_{{\widehat{H}}}\cdot\nabla,J\boldsymbol{\widehat{\Gamma}}\big\rangle
\Big)
\\
=&\, 
\langle \bX_{\widehat{H}}\rangle+\frac1{2\|\Upsilon\|^{2}}\Big(\big\langle (\|\Upsilon\|^{2}J\boldsymbol{\widehat{\Gamma}}\cdot\nabla),\bX_{{\widehat{H}}}\big\rangle
-\big\langle    (\bX_{\widehat{H}}\cdot\nabla),\|\Upsilon\|^{2}J\boldsymbol{\widehat{\Gamma}}\big\rangle
\Big),
\label{Xnewvers}
\end{align}
where
$\boldsymbol{\widehat{\Gamma}}={i\hbar}\|\Upsilon\|^{-2}[\Upsilon\Upsilon^\dagger,\nabla(\Upsilon\Upsilon^\dagger)]=2\boldsymbol{\cal A}_{B}\Upsilon\Upsilon^\dagger+i\hbar\|\Upsilon\|^{-2}(\Upsilon\nabla\Upsilon^\dagger-(\nabla\Upsilon)\Upsilon^\dagger)$
is a non-Abelian gauge potential already appeared in equation \eqref{hannah}.
}

Then, together with \eqref{vectfield}, equation \eqref{NHWE} gives a \emph{nonlinear hybrid wave equation} (NHWE) obtained by a suitable modification of the variational structure underlying the original quantum-classical wave equation \eqref{QCWE}.  The NHWE is very complicated in several ways, due to the cumbersome expression of the phase-space vector field $\boldsymbol{\cal X}$. Thus, suitable numerical methods will need to be developed to approach the associated initial value problem and we are currently pursuing this direction. Here, we will simply notice that, in both the purely quantum and the purely classical cases, one has $\widetilde{\bX}_{\widehat{H}}=0$. Then, in the purely quantum case one recovers the standard Schr\"odinger equation $i\hbar\partial_t\Upsilon=\widehat{H}\Upsilon$, while in the purely classical case, one has $\boldsymbol{\cal X}=\bX_H$ and \eqref{NHWE} returns the KvH equation  $i\hbar
\partial_t\Upsilon=\widehat{\cal L}_H\Upsilon$. {\color{black}Notice that in the present context the  phase term $-(\boldsymbol{\cal A}\cdot\bX_H-H)$ appearing in the prequantum operator $\widehat{\cal L}_H$ is an irrelevant phase factor. Indeed, we need to recall that the classical density is now given by $\rho_c=\|\Upsilon\|^2$, rather than the expression in \eqref{cden}.
}

The \emph{fluctuation vector field} $\widetilde{\bX}_{\widehat{H}}={\bX}_{\widehat{H}}-\langle{\bX}_{\widehat{H}}\rangle$ deserves some words. Notice that $\langle\widetilde{\bX}_{\widehat{H}}\rangle=0$ and if 
\beq\label{simpham}
\widehat{H}(q,p)=\widehat{H}_Q(\hat{x},\hat{p})+H_C(q,p)+\widehat{V}_I(q,\hat{x})
\,,
\eeq
then the expression 
\beq\label{simpvectfield}
\widetilde{\bX}_{\widehat{H}}=(0,\widehat{F}-\langle \widehat{F}\rangle)
\,,\qquad\text{with}\qquad
\widehat{F}=-\partial_q\widehat{V}_I(q,\hat{x})
\,,
\eeq
 identifies the force fluctuation $\widetilde{F}:=\widehat{F}-\langle \widehat{F}\rangle$ from the  Hellmann-Feynman average $\langle \widehat{F}\rangle$ \cite{Feynman2}. 
As we will see later on, neglecting the fluctuation force $\widetilde{F}$ leads to a particularly simple model.

In the remainder of this paper, we will discuss various interesting features arising from the NHWE. Before moving on to this discussion,  we point out that the NHWE may be regarded as a closure scheme based on von Neumann operators and their Wigner transform, similarly to the discussion in Appendix \ref{sec:KoopMixtures}. This is the topic of Appendix \ref{sec:vNOp}.

\section{Discussion\label{sec:discussion}}

This section presents a series of implications of the NHWE model comprised by the equations \eqref{NHWE} and \eqref{vectfield}. As we will see, this model enjoys several properties including the presence of invariants associated to a Hamiltonian structure combining those of quantum and classical mechanics alone. Later on in this section, we will consider how the NHWE model recovers important special cases occurring in the chemical physics literature and the comparison between the full model and its special cases will lead us to make considerations on the presence of the quantum backreaction on the classical trajectories. At first, we will present how Poincar\'e's integral invariant transfers to the quantum-classical setting.

\subsection{Poincar\'e integral invariant\label{sec:PIinv}}
The NHWE model discloses an interesting feature: as a consequence of the first line in \eqref{patrizia}, we obtain the hybrid Poincar\'e integral invariant
\beq\label{PII}
\frac{\de}{\de t}\oint_{\boldsymbol{c}(t)}\left(p\de q-\boldsymbol{\cal A}_B(\bz,t)\cdot\de\bz\right)=
0
\,,
\eeq
where $\boldsymbol{c}(t)=\boldsymbol{\eta}(\boldsymbol{c}_0,t)$ and $\boldsymbol{c}_0$ is an arbitrary loop in the classical phase-space. 
In terms of differential forms, one may introduce the Berry curvature 
\[
{\cal B}=
\nabla\boldsymbol{\cal A}_B-(\nabla\boldsymbol{\cal A}_B)^T
=\frac{2\hbar}{\|\Upsilon\|^2}\big(\operatorname{Im}\langle\nabla\Upsilon|\nabla\Upsilon\rangle-\operatorname{Im}\langle\Upsilon|\nabla\Upsilon\rangle\wedge\operatorname{Re}\langle\Upsilon|\nabla\Upsilon\rangle\big)
\,,
\]
where we have used the wedge product notation $\bv\wedge\bw=\bv\bw-\bw\bv$.
Then, applying Stokes theorem to \eqref{PII} yields the following evolution for the exact two-form $\Omega(t)=J-{\cal B}(t)$:
\beq
\Omega_{jk}(\boldsymbol{\eta}(\bz_0,t),t)\, \de {\eta}^j(\bz_0,t)\wedge\de {\eta}^k(\bz_0,t)=\Omega_{0\, jk}(\bz_0)\,\de z_0^j\wedge\de z_0^k
\,,
\eeq
which is analogous to the relation for Hamiltonian flows in symplectic geometry. However, in this case the differential 2-form $\Omega(t)$ is time-dependent and it remains symplectic at all times if it is so initially.

Notice that, by the usual identification between maximal forms and densities,  the Liouville theorem ensures that the density $\operatorname{\sf det}\Omega$ satisfies the continuity equation $\partial_t(\operatorname{\sf det}\Omega)+\operatorname{div}(\boldsymbol{\cal X}\operatorname{\sf det}\Omega)=0$. Consequently, any one-variable function $\Phi({\sf x})$ generates the invariant functional
\[
{\cal I}(\Upsilon)=\int\! \|\Upsilon\|^{2\,}\Phi\bigg(\frac{\operatorname{\sf det}\Omega}{\|\Upsilon\|^2}\bigg)\,\de^2z
\,.
\]
For example, upon recalling \eqref{gigi} and its preceding lines, we observe that in the purely classical case the Berry connection is a pure differential so that the invariant $\int \|\Upsilon\|^{2}\ln({\operatorname{\sf det}\Omega}/{\|\Upsilon\|^2})\,\de^2z$ recovers the usual expression of Gibbs' entropy. The identification of entropy functionals in the context of quantum-classical dynamics is an interesting question that is currently under investigation.

\rem{ 
Upon using the second line in \eqref{patrizia} and recalling the first line of \eqref{vectfield}, the Poincar\'e invariant relation \eqref{PII} may be rewritten in the non-conservative form as
\beq
\frac{\de}{\de t}\oint_{\boldsymbol{c}(t)}p\de q=\oint_{\boldsymbol{c}(t)}J\boldsymbol{\cal X}\cdot\de\bz
\,.
\eeq
In the absence of coupling one has $\widehat{H}(\bz)=H_C(\bz)+\widehat{H}_Q$,  so that $J\boldsymbol{\cal X}\cdot\de\bz=-\de H_C$ and the relation above recovers the classical Poincar\'e integral invariant $\oint p\de q=const$.
It may be interesting to notice the substantial difference between the previous relation and the corresponding relation that emerges from the original hybrid model, before applying the closure relation \eqref{closrel}. Indeed, we recall that the Euler-Poincar\'e Lagrangian \eqref{Tiziana2} leads immediately to $\boldsymbol{\cal X}=\langle\bX_{\widehat{H}}\rangle$ and, upon using $\langle\bX_{\widehat{H}}\rangle\cdot\nabla\boldsymbol{\cal A}
+\nabla\langle\bX_{\widehat{H}}\rangle\cdot\boldsymbol{\cal A}
=\nabla(\langle\bX_{\widehat{H}}\rangle\cdot\boldsymbol{\cal A})+J\langle \bX_{\widehat{H}}\rangle$, we have
\beq\label{EhrenfestPIEvol}
\frac{\de}{\de t}\oint_{\boldsymbol{c}(t)}p\de q=\oint_{\boldsymbol{c}(t)}J\langle \bX_{\widehat{H}}\rangle\cdot\de\bz=-\oint_{\boldsymbol{c}(t)}\langle\nabla{\widehat{H}}\rangle\cdot\de\bz
\eeq
for any loop $\boldsymbol{c}(t)$ moving with the phase-space vector field $\langle \bX_{\widehat{H}}\rangle$.
}  

The existence of a Poincar\'e integral invariant unfolds some of the symplectic properties of the NHWE \eqref{NHWE}. As recently shown in \cite{GBTr21} by using the density matrix formalism, the closure model provided by the NHWE possesses a very rich geometric content and here we continue this discussion by presenting the Hamiltonian structure underlying equations  \eqref{NHWE} and  \eqref{vectfield}.

\color{black}
\subsection{Density matrix and Hamiltonian structure\label{sec:GIHS}}

As the NHWE model is derived from the variational principle associated to the Lagrangian \eqref{untangledLagr}, it is expected that its equations of motion are Hamiltonian with respect to the Hamiltonian functional \eqref{QCHam}. In order to present the Hamiltonian structure, it is convenient to rewrite the NHWE \eqref{prelimNHWE} in terms of the density matrix $\widehat{\cal P}(\bz)=\Upsilon(\bz)\Upsilon^\dagger(\bz)$. This is an easy step upon recognizing that the Hamiltonian functional \eqref{QCHam} may be rewritten as
\beq\label{QCHamP}
h_{\scriptscriptstyle QC}( \widehat{\cal P})=
\!
\int\! \bigg\langle \widehat{\cal P},\widehat{H}
+\frac{i\hbar}{2\operatorname{\sf Tr}\widehat{\cal P}}\, \big[\nabla\widehat{\cal P},\bX_{\widehat{H}}\big]\bigg\rangle\,\de^2z
\eeq
and by resorting to the chain rule, that is $\delta h_{\scriptscriptstyle QC}/\delta \Upsilon=2(\delta h_{\scriptscriptstyle QC}/\delta\widehat{\cal P}) \Upsilon$. These relations allow rewriting  \eqref{prelimNHWE} as $i\hbar\partial_t\widehat{\cal P}+i\hbar\operatorname{div}(\boldsymbol{\cal X}\widehat{\cal P})=[\delta h_{\scriptscriptstyle QC}/\delta\widehat{\cal P},\widehat{\cal P}]$. In addition, the vector field $\boldsymbol{\cal X}$ may be written in terms of $\widehat{\cal P}$ by using the chain rule in the first line of \eqref{vectfield}. One finds $\boldsymbol{\cal X}=J\langle\widehat{\cal P},\nabla(\delta h_{\scriptscriptstyle QC}/\delta\widehat{\cal P})\rangle/\operatorname{\sf Tr}\widehat{\cal P}=\langle\bX_{\delta h_{\scriptscriptstyle QC}/\delta\widehat{\cal P}}\rangle$, where we have used $\langle\widehat{A}\rangle=\langle\widehat{\cal P},\widehat{A}\rangle/\operatorname{\sf Tr}\widehat{\cal P}$. In conclusion, one obtains  equation \eqref{andrea} with $\widehat{\mathcal{H}}=\delta h/\delta\widehat{\cal P}$ and
\beq\label{gnrtr}
\frac{\delta h_{\scriptscriptstyle QC}}{\delta\widehat{\cal P}}= \widehat{H}+\frac1{2\operatorname{\sf Tr}\widehat{\cal P}}\bigg(2{i\hbar}\big(\{\widehat{\cal P},{\widehat{H}}\}+\{\widehat{H},\widehat{\cal P}\}\big)+{i\hbar}[\{\ln \operatorname{\sf Tr}\widehat{\cal P},\widehat{H}\},\widehat{\cal P}] -
\big\langle i\hbar\{\widehat{\cal P},{\widehat{H}}\}+i\hbar\{\widehat{H},\widehat{\cal P}\}\big\rangle \boldsymbol{1}\bigg).
\eeq
Notice that equation \eqref{andrea} is gauge-independent and thus the pure phase term $-\boldsymbol{\cal A}\cdot\boldsymbol{\cal X}\Upsilon$ in \eqref{prelimNHWE} plays no role in this setting. In addition, upon using $\langle \bX_{\,\widehat{\mathcal{H}}}\rangle
=
\langle\rho, \bX_{\, \widehat{\mathcal{H}}}\rangle
=
\bX_{\langle\rho,  \widehat{\mathcal{H}}\rangle}-\langle \bX_\rho,\! \widehat{\mathcal{H}}\rangle$, with $\rho=\widehat{\cal P}/\operatorname{Tr}{\widehat{\cal P}}$,  the Leibniz product rule yields the  expression of $\boldsymbol{\cal X}$ in \eqref{Xnewvers} where $\boldsymbol{\widehat{\Gamma}}={i\hbar}[\rho,\nabla\rho]$.

Equation \eqref{andrea} was shown to be Hamiltonian in \cite{GBTr21}. In particular, upon introducing the notation $A:B= \operatorname{\sf Tr}(AB)$, its Poisson bracket reads
\begin{equation}\label{bracket_candidate_rho}
\{k,h\}_{\widehat{\cal P}}=-\int  \!\left\langle \widehat{\cal P}  ,\frac{i}\hbar\!\left[\frac{\delta k}{\delta \widehat{\cal P}},\frac{\delta h}{\delta \widehat{\cal P}}\right] \right\rangle\de^2z +\int \!\frac1{\operatorname{\sf Tr}\widehat{\cal P}  }\bigg(\widehat{\cal P}  : \left\{\frac{\delta k}{\delta \widehat{\cal P}},\frac{\delta h}{\delta \widehat{\cal P}}\right\}: \widehat{\cal P}   \bigg) \,\de^2z \,,
\end{equation} 
and this is accompanied by the Hamiltonian functional \eqref{QCHamP}. A thorough discussion of the Poisson bracket \eqref{bracket_candidate_rho} was given in \cite{GBTr21}. The bracket structure for the NHWE is found from \eqref{bracket_candidate_rho}  by simply using the chain rule. One obtains
\beq
\{k,h\}_\Upsilon=\int\!\bigg(\underbrace{\frac1{2\hbar}\operatorname{Im}\left\langle\frac{\delta k}{\delta\Upsilon}\bigg|\frac{\delta h}{\delta\Upsilon}\right\rangle}_\textit{\tiny Quantum Schr\"odinger bracket}+\underbrace{\frac1{\|\Upsilon\|^2}\bigg(\Upsilon\diamond\frac{\delta k}{\delta \Upsilon}\bigg)\cdot J\bigg(\Upsilon\diamond\frac{\delta h}{\delta \Upsilon}\bigg)}_\textit{\tiny Classical Koopman-von Neumann bracket}
\bigg)\de^2 z
\,.
\label{PB}
\eeq
where
\[
\Upsilon\diamond\frac{\delta h}{\delta \Upsilon}:=\frac1{2\|\Upsilon\|^2}\bigg(\bigg\langle\frac{\delta h}{\delta \Upsilon},\nabla\Upsilon\bigg\rangle-\bigg\langle\Upsilon,\nabla\frac{\delta h}{\delta \Upsilon}\bigg\rangle\bigg)\,.
\]
Importantly, due to its construction, this bracket structure  becomes  Poisson  when restricted to gauge-invariant functionals, which is indeed the case for \eqref{QCHam}.
Notice that the bracket \eqref{PB} returns equation \eqref{NHWE} in the gauge  obtained by dropping the pure phase term $-\boldsymbol{\cal A}\cdot\boldsymbol{\cal X}\Upsilon$. 

The  bracket \eqref{PB} has a particularly suggestive form in that it is the sum of the canonical quantum bracket for the evolution of Schr\"odinger wavefunctions and the classical KvN bracket  \eqref{KvNHS2}  from Section \ref{KvNHS2}. For earlier attempts to construct Poisson brackets in quantum-classical dynamics, we address the reader to \cite{Prezhdo,PrKi}.

\color{black}

\subsection{Hybrid density operator\label{sec:HybDenOp}}
In this section we are interested in the analogue of the hybrid density operator \eqref{hybden} in the context of the NHWE model. Upon performing a similar integration by parts to the second equality in \eqref{toten}, we observe that the NHWE Hamiltonian functional \eqref{QCHam} may be rewritten as
\beq
\int\langle\Upsilon,\widehat{\cal L}_{\widehat{H}}\Upsilon+(\boldsymbol{\cal A}-\boldsymbol{\cal A}_B)\cdot \bX_{\widehat{H}}\Upsilon
\rangle\,\de^2z=\operatorname{\sf Tr}\int\!\widehat{\cal D}\widehat{H}\,\de^2z
\,,
\eeq
where
\beq\label{hybden2}
\widehat{\cal D}:=\Upsilon\Upsilon^\dagger-\operatorname{div}(J\boldsymbol{\cal A}_B\Upsilon\Upsilon^\dagger)+i\hbar\{\Upsilon,\Upsilon^\dagger\}
\eeq
and all quantities are evaluated at the same phase-space point $\bz=(q,p)$. Notice that the original expression in \eqref{hybden} is obtained by simply replacing $\boldsymbol{\cal A}_B\to\boldsymbol{\cal A}$ in \eqref{hybden2}. As we showed in \cite{GBTr21}, the hybrid operator \eqref{hybden2} may also be written in terms of the variable $\color{black}\widehat{\cal P}=\Upsilon\Upsilon^\dagger$. Similarly to \eqref{hybden}, the new density operator is sign-indefinite while it consistently recovers a  positive quantum density matrix $\hat\rho_q=\int\!\widehat{\cal D}\,\de^2z$ and a positive classical density $\rho_c=\operatorname{\sf Tr}\widehat{\cal D}$. We recall from the original treatment in Section \ref{sec:hybrids} that the expression \eqref{cden} was not positive-definite. 

More importantly, in \cite{GBTr21} we proved that the hybrid operator \eqref{hybden2} is covariant with respect to both canonical transformations $\boldsymbol\upeta$ in phase-space and unitary transformations ${\cal U}$ of the quantum state space. Upon denoting the functional dependence of $\widehat{\cal D}(\bz)$ on $\Upsilon$ by $\widehat{\cal D}[\Upsilon(\bz)]$, we have the properties
\[
\widehat{\cal D}[\Upsilon( {\boldsymbol\upeta}(\bz))]= \widehat{\cal D}[\Upsilon] ( {\boldsymbol\upeta}(\bz))
\qquad\text{and}\qquad
\widehat{\cal D}[{\cal U}\Upsilon]={\cal U}\widehat{\cal D}[\Upsilon]{\cal U}^\dagger.
\]
Here, ${\cal U}$ is a quantum unitary operator while $\boldsymbol\upeta$ is a classical canonical transformation, so that $J_{jk}\, \de {\upeta}^j(\bz)\wedge\de {\upeta}^k(\bz)=J_{jk}\,\de z^j\wedge\de z^k$. As showed in \cite{GBTr21}, these covariance properties lead to casting the equations for the classical density $\rho_c=\|\Upsilon\|^2$ and the quantum density matrix $\hat\rho_q=\int\! \Upsilon\Upsilon^\dagger\de^2z$ as
\beq\label{densevol2}
\frac{\partial\rho_c}{\partial t}=\operatorname{\sf Tr}\{\widehat{H},\widehat{\cal D}\}
\,,\qquad\qquad
i\hbar\frac{\de\hat\rho_q}{\de t}=\int[\widehat{H},\widehat{\cal D}]\,\de^2z
\,,
\eeq
which are formally identical to the analogous relations \eqref{classevol} and \eqref{quantevol} in the original treatment.
The relations \eqref{densevol2} govern the Ehrenfest equations for quantum and classical expectation values. In general, we define the expectation value  of a hybrid observable $\widehat{A}=\widehat{A}(\bz)$ as $\langle\widehat{A}\rangle_\Upsilon:=\operatorname{\sf Tr}\!\int\!\widehat{\cal D}\widehat{A}\,\de^2z$. Then, if $\widehat{A}={A}_C(\bz)$ is a classical observable and $\widehat{A}_Q$ is a quantum observable (independent of $\bz$), their Ehrenfest equations are
\[
\frac{\de}{\de t}\langle A_C\rangle_\Upsilon=\langle \{A_C,\widehat{H}\}\rangle_\Upsilon
\,,\qquad\quad\text{and}\qquad\quad
i\hbar\frac{\de}{\de t}\langle \widehat{A}_Q\rangle_\Upsilon=\langle [\widehat{A}_Q,\widehat{H}]\rangle_\Upsilon
\,.
\]
While no simple Ehrenfest equation is available for general hybrid observables, the relations above provide already much insight.
For example, it is easy to see that hybrid Hamiltonians of the type $\widehat{H}=m^{-1}\hat{p}^2/2+M^{-1}p^2/2+\widehat{V}(q-\hat{x})$ yield conservation of that total momentum, that is
${\de\langle p+\hat{p}\rangle_\Upsilon}/\de t=\langle \{p,\widehat{V}\}-i\hbar^{-1} [\hat{p},\widehat{V}]\rangle_\Upsilon=0$.

 \subsection{Specializations and further comments\label{sec:SpecsFC}}

This section is devoted to presenting relevant special cases of the NHWE model in quantum-classical dynamics. As we already discussed after equation \eqref{vectfield}, the purely quantum and purely classical cases are immediately recovered by the present treatment. Here, we will show how simple quantum-classical models are also naturally recovered. Eventually, we will conclude this section by making considerations on the appearance of the quantum backreaction on the classical Lagrangian paths.

\medskip
\noindent{\bf Mean-field model.} If $\Upsilon(\bz,x)=\psi(x)\sqrt{D(\bz)}e^{iS(\bz)/\hbar}$ (with $ \psi $ independent of $\bz$), then the NHWE Hamiltonian \eqref{QCHam} reduces to $h=\int \!D\langle\widehat{H}\rangle\,\de^2z$ and $\boldsymbol{\cal A}_B=\nabla S$. Thus, the term $\langle\Upsilon,\boldsymbol{\cal A}_B\cdot\boldsymbol{\cal X}\Upsilon\rangle$ reduces to a pure time derivative and the Lagrangian \eqref{untangledLagr} drops to
$\ell_{\rm MF}=\int\! D\big(\boldsymbol{\cal A}\cdot\boldsymbol{\cal X}+\langle\psi,i\hbar \dot\psi-\widehat{H}\rangle 
\big)\,\de^2z$,
whose associated variational principle returns the mean-field equations \eqref{MFeqs} for $\rho_c=D$ and $\hat\rho_q=\psi\psi^\dagger$.

\medskip
\noindent{\bf Ehrenfest model.} 
If we assume $\widetilde\bX_{\widehat{H}}\simeq0$ and discard the corresponding term in the NHWE Hamiltonian \eqref{QCHam}, then the latter reduces to $h(\Upsilon)=\langle\widehat{H}\rangle_\Upsilon$. In this case, we have $\boldsymbol{\cal X}=\langle\bX_{\widehat{H}}\rangle$ and the NHWE \eqref{NHWE} reduces to 
\beq
i\hbar\frac{\partial \Upsilon}{\partial t}+i\hbar\langle\bX_{\widehat{H}}\rangle\cdot\nabla\Upsilon+\frac{i\hbar}2\Upsilon\operatorname{div}\langle\bX_{\widehat{H}}\rangle={\widehat{H}}
\Upsilon\color{black}-\langle p\partial_p\widehat{H}\rangle\Upsilon
\,,
\label{Ehrenfest}
\eeq
 In terms of the measure-valued density matrix $\widehat{\cal P}=\Upsilon\Upsilon^\dagger$, the hybrid wave equation  \eqref{Ehrenfest} reads
$
{i\hbar{\partial_t \widehat{\cal P}}}+{i\hbar\operatorname{div}(\widehat{\cal P}\langle\bX_{\widehat{H}}\rangle)}=[{\widehat{H}}
,\widehat{\cal P}]
\,.
$
Recently, this model was also obtained  as a closure model directly from the original theory in Section \ref{sec:hybrids}; see \cite{GBTr21b}. 
The equations for the classical density $\rho_c=\operatorname{\sf Tr}\widehat{\cal P}$ and the quantum density matrix $\hat\rho_q=\int\widehat{\cal P} \de^2z$ are
\beq
\frac{\partial \rho_c}{\partial t}+\operatorname{div}(\rho_c\langle\bX_{\widehat{H}}\rangle)=0
\,,\qquad\qquad\ 
i\hbar\frac{\de {\hat\rho}_q}{\de t}=\int[\widehat{H},\widehat{\cal P}]\,\de^2z
\,,
\label{Ehrenfest2}
\eeq
respectively.
We notice that, unlike the mean-field model \eqref{MFeqs},  the present construction allows to capture some decoherence effects. Indeed, the second in \eqref{Ehrenfest2} implies that  the quantum purity $\|\hat\rho_q\|^2=\operatorname{\sf Tr}(\hat\rho_q^2)$ possesses nontrivial dynamics. Similar conclusions were reached previously in \cite{Alonso}, although in most cases the decohrence levels obtained  by this model -- usually known as the \emph{Ehrenfest model} -- are considerably lower than those obtained from fully quantum predictions. {\color{black}Another limitation of Ehrenfest dynamics arises in the context of pure-dephasing systems, that is hybrid systems involving a quantum two-level subsystem and whose hybrid Hamiltonian is $\widehat{H}=H_c+H_I\widehat{\sigma}_k$, where $\widehat{\sigma}_k$ is any of the three Pauli matrices. In this case, Ehrenfest dynamics gives  $\partial_t\langle\widehat{\sigma}_k\rangle+\langle\bX_{\widehat{H}}\rangle\cdot\nabla\langle\widehat{\sigma}_k\rangle=0$ with $\langle\bX_{\widehat{H}}\rangle=\bX_{{H}_c}+\bX_{{H}_I}\langle\widehat{\sigma}_k\rangle$. Then, if $\langle\widehat{\sigma}_k\rangle=0$ initially, this relation holds indefinitely and the classical evolution in \eqref{Ehrenfest2} returns simply ${\partial_t\rho_c}=\{H_c,\rho_c\}$. Thus, we observe that the classical dynamics decouples completely from the quantum motion. This quantum-classical decoupling occurring for pure-dephasing systems is among the main drawbacks of the Ehrenfest model. Instead, this issue is absent in the NHWE, or its density-matrix formulation given by \eqref{andrea}, with $\widehat{\cal H}=\delta h/\delta \widehat{\cal P}$ and \eqref{gnrtr}. Indeed, in this case one obtains $ (\partial_t+\boldsymbol{\cal X}\cdot\nabla )\langle\widehat{\sigma}_k\rangle= D^{-1}\{H_I, D(\langle\widehat{\sigma}_k\rangle^2-1)\}$, with $\boldsymbol{\cal X}=\bX_{H_c}+\bX_{H_I}\langle\widehat{\sigma}_k\rangle+
\rho_c^{-1}  \big( 
\big\langle\rho_c J\boldsymbol{\widehat{\Gamma}},\widehat{\sigma}_k\big\rangle\cdot \nabla\bX_{H_I}
-
 \bX_{{H}_I}\cdot\nabla\big\langle\rho_c J\boldsymbol{\widehat{\Gamma}},\widehat{\sigma}_k\big\rangle
\big)
/2$ and $\rho_c\boldsymbol{\widehat{\Gamma}}=i\hbar[\widehat{\cal P},\nabla\widehat{\cal P}]$. See \cite{GBTr21} for more discussions on how the closure model proposed here adapts to the case of quantum two-level systems.
}

\medskip
\noindent{\bf Quantum decoherence and backreaction force.} While quantum decoherence is absent in the mean-field model \eqref{MFeqs} and is poorly captured by the  Ehrenfest wave equation \eqref{Ehrenfest},  the full NHWE model \eqref{NHWE} seems to suggest that decoherence and other quantum-classical correlation effects arise from the fluctuation vector field $\widetilde{\bX}_{\widehat{H}}$, which is given  by \eqref{simpvectfield} for the standard-type Hamiltonian operator \eqref{simpham}.
Then, the  fluctuation force $\widetilde{F}=\widehat{F}-\langle\widehat{F}\rangle$ around the Hellmann-Feynman average $\langle\widehat{F}\rangle$ emerges in the present framework as the main responsible for quantum-classical correlations. Besides quantum decoherence, this Hellmann-Feynman  fluctuation  $\widetilde{F}$ also comprises the so-called {\it quantum backreaction}, that is the feedback force on the classical trajectories. While the mean-field vector field $\bX_{\langle\widehat{H}\rangle}$ occurring in the first equation $\partial_t\rho_c+\operatorname{div}(\rho_c\bX_{\langle\widehat{H}\rangle})=0$ of  \eqref{MFeqs} ignores this important effect,  part of the backreaction appears  in the Ehrenfest vector field $\langle\bX_{\widehat{H}}\rangle=\bX_{\langle\widehat{H}\rangle}+(\langle\bX_{\widehat{H}}\rangle-\bX_{\langle\widehat{H}\rangle})$ from the first  equation in \eqref{Ehrenfest2}. Unfortunately, the Ehrenfest model is known to lack most backreaction effects \cite{Tully}. In the more general case of equation \eqref{vectfield}, however, we observe that  the rest of the backreaction effects is retained by the  difference $\boldsymbol{\cal X}-\langle\bX_{\widehat{H}}\rangle$, which for the Hamiltonian \eqref{simpham} is   triggered entirely by the  fluctuation force in ${\bX}_{\widehat{H}}-\langle\bX_{\widehat{H}}\rangle=(0,\widetilde{F})$. Then, the quantity $\widetilde{F}$ emerges as a fundamental object in quantum-classical coupling and we shall call it the \emph{backreaction force}.

\section{Summary and outlook}

Based on Sudarshan's early attempt in \cite{Sudarshan}, this paper has formulated a dynamical theory of hybrid quantum-classical systems by following two sequential steps. In the first, we followed our original work by blending the variational structure of Koopman and Schr\"odinger wavefunctions after exploiting the KvH construction to retain classical phases within the formalism. In the second, we took a step beyond our original model by making classical phases unobservable via a gauge-invariance principle. As we showed, the second step is crucial in ensuring that the classical state is always identified by a positive-definite Liouville density on phase-space, whose transport equation is now directly related to Noether's theorem. While classical phases were made unobservable by enforcing superselection rules following Sudarshan's work, the latter led to substantial issues which are here circumvented by resorting to a simpler gauge principle. The idea of a gauge principle reflects previous comments made by Ghose \cite{Ghose}, and much earlier by Boucher and Traschen \cite{boucher}, on the role of classical phases in hybrid dynamics.

As we showed, the introduction of a gauge principle in the original variational principle requires nontrivial sophisticated steps in geometric mechanics combining tools from continuum dynamics, such as Lagrangian paths and Lie transport, with Hilbert-space methods from quantum and Koopman classical mechanics, e.g. phase-space wavefunctions and local density matrices. In particular, a key ingredient is the exact factorization method allowing the identification of the classical phase and ultimately leading to the application of the gauge principle.

The final NHWE model comprised by \eqref{NHWE} and \eqref{vectfield} consists of a formidable equation for a hybrid QC wavefunction. Unlike the original formulation in \eqref{QCWE}, the  NHWE equation is made highly nonlinear by the presence of the Berry connection in the Lagrangian \eqref{untangledLagr}. Yet, in this context the Berry connection is crucial in realizing the QC correlations. Given the level of complexity of the NHWE model,  {\color{black}trajectory-based computational tools analogue to those in \cite{FoHoTr19,HoRaTr21} are currently under development in order to account for several degrees of freedom and strong subsystem-bath couplings.} Also, further closure models are desirable to alleviate the difficulties arising from a full phase-space treatment. At present, fluid closure models are  under investigation.

Despite its complexity, the NHWE model was shown to retain several remarkable features that are not easy to find in alternative hybrid theories. For example, the model possesses a hybrid Poincar\'e integral invariant which is unavailable in the original formulation. As showed, this leads to defining a time-dependent hybrid symplectic form whose associated Liouville volume can be used to construct dynamical invariants. Indeed, the latter are allowed in the form of Casimir functionals by the presence of a noncanonical Hamiltonian structure that is written in \eqref{PB} as the sum of a quantum Schr\"odinger term and a classical Koopman term. In turn, the existence of Casimir invariants allows  the identification of Lyapunov-stable equilibria by following the methods in \cite{HoMaRaWe85}. 
{\color{black}In addition, as discussed in the previous section, the NHWE overcomes the quantum-classical decoupling occurring in pure-dephasing Ehrenfest dynamics, thereby ensuring the persistence of a quantum backreaction on the classical evolution.
}

We conclude by emphasizing two more different directions arising from this work. At a foundational level, {\color{black}we would like to understand how the results in this paper can be used to formulate a dynamical theory of quantum measurement. As pointed out by Peres \cite{Peres}, the latter can be envisioned as the combination of a reversible \emph{pre-measurement}  mechanism and an irreversible process. The latter is due to the fact that the classical apparatus is realistically in a thermal equilibrium. This description requires the addition of entropy sources in the classical sector, thereby indicating a possible way forward in the formulation of a measurement theory.}

On a more applied level, we would also like to apply the present theory to devise new models for quantum-classical spin hybrids \cite{RuKaUp22}. The general idea is to reach previously unaccessible regions in the control parameter space of spin control protocols  by exploiting the coupling to large dipole moments of classical magnets. This picture involves the interaction dynamics of a classical magnetization vector with quantum spins, in which case the former obeys  Landau-Lifshitz-type dynamics. We plan to pursue this interesting direction in future work.

\paragraph{Acknowledgments.}  
We are grateful to Andr\'es Berm\'udez, Francesco Di Maiolo, Darryl Holm, Ilon Joseph, Phil Morrison, and Robert MacKay for their keen remarks during the development of this work. 
This work was made possible through the support of Grant 62210 from the John Templeton Foundation. The opinions expressed in this publication are those of the authors and do not necessarily reflect the views of the John Templeton Foundation. Partial support by the Royal Society Grant IES\textbackslash R3\textbackslash203005 is also greatly acknowledged. 

\addtocontents{toc}{\protect\setcounter{tocdepth}{0}}
\appendix

\section{Von Neumann operators in Koopman  mechanics\label{sec:KoopMixtures}}

This appendix  presents an extension of the treatment in Section \ref{sec:BackToKvN} that overcomes the difficulties arising from the use of the relation $\nabla S=\boldsymbol{\cal A}$. As  discussed in various instances, this relation requires a singular phase which may pose several problems. Here, we will overcome this difficulty by extending the  wavefunction treatment from Section \ref{sec:BackToKvN} to consider the more general setting made available by  von Neumann operators and their Wigner transforms.

Let us start from KvH theory. Instead of considering the KvH equation \eqref{KvH2} for the wavefunction $\chi(\bz,t)$, here we consider the Koopman mixture
\beq
\widehat{\Theta}(\bz,\bz')=\sum_{a=1}^Nw_a\chi_a(\bz)\chi_a^*(\bz')
\,,
\label{vnoper}
\eeq
with $w_a\geq0$ and $\sum_{a=1}^Nw_a=1$. The von Neumann operator $\widehat{\Theta}$ mimics the density matrix from standard quantum theory and satisfies the equation
\beq
i\hbar\partial_t\widehat{\Theta}=[\widehat{\cal L}_H,\widehat{\Theta}]
\,,\qquad
\text{where}
\qquad
\widehat{\cal L}_H:=i\hbar\{ H,\ \}+\mathscr{L}
\label{vneqn}
\eeq
is the covariant Liouvillian from \eqref{KvH2}.  Then, the variational structure for the Koopman mixture is given by the natural extension of the KvH Lagrangian $L_{\rm KvH}$ in \eqref{VP-KvH}, that is
\beq
L_{\rm mix}=\sum_{a=1}^Nw_aL_{\rm KvH}(\chi_a,\partial_t\chi_a)
\,,
\label{VP-mKvH}
\eeq
where the subscript `mix' refers to the fact that we are dealing with a mixture. This Lagrangian yields a sequence of identical KvH equations for each $\chi_a$, with the  same Hamiltonian $H$. Then, we can write $\chi_a=\sqrt{D_a}e^{iS_a/\hbar}$ and perform exactly the same steps in the previous section. For example, the Lagrange-to-Euler map \eqref{LtoEmap} now reads $D_a(\bz,t)=\int\!D_{a,0}(\bz)\delta(\bz-\boldsymbol{\eta}(\bz_0,t))\,\de^2z_0$, or equivalently in terms of the Jacobian  ${\cal J}_{\boldsymbol\eta}=\operatorname{\sf det}\nabla\boldsymbol{\eta}$. We note that, since the $\chi_a$'s all obey the same KvH equation, the Lagrangian path $\boldsymbol{\eta}$ is the same  for each $a=1,...,N$. Upon defining
\beq
D=\sum_{a=1}^Nw_aD_a
\,,\qquad\text{ and }\qquad
\bsigma=\sum_{a=1}^Nw_aD_a\nabla S_a
\,,
\label{momaps}
\eeq
the classical Liouville density reads
\beq
\label{mario}
\rho_c=D+\operatorname{div}(J(\bsigma-D\boldsymbol{\cal A}))
\eeq
and the Lagrangian \eqref{VP-mKvH} is taken into the Euler-Poincar\'e form
\beq
L_{\rm EPmix}=\int\!(\bsigma\cdot\boldsymbol{\cal X}+(D\boldsymbol{\cal A}-\bsigma)\cdot J\nabla H- DH)\,\de^2z
\,,
\label{EP-mKvH}
\eeq
which replaces the previous expression in \eqref{EP-KvH}.
It may be worth noticing that the quantities \eqref{momaps} may also be written entirely in terms of the operator $\widehat{\Theta}$ as \cite{FoHoTr19}
\beq\label{momaps2}
D(\bz)=\widehat\Theta(\bz,\bz')|_{\bz'=\bz}
\,,\qquad\text{ and }\qquad
\bsigma(\bz)=\frac12\big[\hat{\blambda},\widehat\Theta\big]_+(\bz,\bz')|_{\bz'=\bz}
\,,
\eeq
{\color{black}where  $[\cdot ,\cdot]_+$ denotes the anticommutator and we have introduced the operator $\hat{\blambda}=-i\hbar\nabla$, so that $[\hat{\boldsymbol{z}},\hat\blambda]=i\hbar\boldsymbol{1}$}.
Given the appearance of the above quantities in \eqref{mario}, here we remark that  the KvH formulation generally allows for a sign-indefinite  operator $\widehat{\Theta}$,  as long as the corresponding classical distribution $\rho_c$ is positive. Indeed, as we will see below, it is convenient to consider von Neumann operators beyond the mixture type \eqref{vnoper}, which however represents a useful tool to obtain the Lagrangian $L_{\rm EPmix}$. 

In the variational formulation associated to \eqref{EP-mKvH}, the variation $\delta\bsigma$ is arbitrary and one makes use of the variations \eqref{Dvar}-\eqref{Xvar}, along with \eqref{D-eq}. Equivalently, one may also write $\bsigma=D\bLambda$ with arbitrary $\delta\bLambda$.  Then, Hamilton's principle $\delta\int_{t_1}^{t_2}L_{\rm EPmix}\,\de t=0$ yields the analogous of \eqref{dS-A}, that is, upon writing $\bLambda:=\bsigma/D$,
\beq
\partial_t(\bLambda-\boldsymbol{\cal A})+\bX_H\cdot\nabla(\bLambda-\boldsymbol{\cal A})+\nabla \bX_H\cdot(\bLambda-\boldsymbol{\cal A})=0
\,.
\label{lambda-A}
\eeq
We realize that, since $\bLambda$ is no longer an exact differential, the preserved condition $\bLambda=\boldsymbol{\cal A}$ may be set at the initial time without incurring in  topological singularities. In this way, the classical density \eqref{mario} collapses to $\rho_c=D$, which again satisfies $\partial_tD=\{H,D\}$. But what does the relation $\bLambda=\boldsymbol{\cal A}$ correspond to in terms of the von Neumann operator $\widehat\Theta$ in \eqref{vneqn}? Can we obtain the relation $\bLambda=\boldsymbol{\cal A}$ as a \emph{closure model} expressing $\bsigma=D\bLambda$ as a function $\bsigma=\bsigma(D,\boldsymbol{\cal A})$? In order to address these points, it is convenient to use Wigner functions.

Let us consider the Wigner transform of equation \eqref{vneqn}, that is
\beq
\label{WMeq}
\partial_t W(\bz,\blambda,t)=\{\!\!\{{\cal L}_H(\bz,\blambda),W(\bz,\blambda,t)\}\!\!\}
\,,\qquad\text{ where } \qquad
{\cal L}_H(\bz,\blambda)=\bX_H(\bz)\cdot\blambda-\mathscr{L}(\bz)
\eeq
is the Weyl symbol of the covariant Liouvillian in \eqref{vneqn}, $W(\bz,\blambda,t)$ is the Wigner transform of $\widehat{\Theta}(\bz,\bz',t)$, and $\{\!\!\{\, ,\, \}\!\!\}$ is the Moyal bracket in the double phase-space coordinates $(\bz,\blambda)$. We notice that, since ${\cal L}_H(\bz,\blambda)$ is linear in $\blambda$, the moment hierarchy $\int\!\blambda^{n\,} W\,\de^2 z$ closes at any order greater than or equal to 1. This can be seen explicitly by taking the $L^2-$pairing of   \eqref{WMeq} with the arbitrary analytic  function $\blambda^n\Phi_n(\bz)$, and then using the properties of the Moyal bracket. 
Thus, we may consider the closed dynamics for the variables given by $n=0,1$, that is  $D=\int\! W\,\de^2 z$ and  $\bsigma=\int\!\blambda^{\,} W\,\de^2 z$, respectively. In particular, we notice that any local Gaussian function of the type
$W(\bz,\blambda,t)={(2\pi\Sigma^2)^{-1}D(\bz,t)}\,\exp\!\big(-{\Sigma^{-2}{\left|\blambda-\bLambda(\bz,t)\right|^2}}/2\big)$
is an exact solution of \eqref{WMeq} for any $\Sigma$, provided that $\bLambda=\bsigma/D$ satisfies \eqref{lambda-A} and $D$ satisfies the classical Liouville equation. 
Thus, setting $\bLambda=\boldsymbol{\cal A}$ we obtain that the Wigner function
\beq
W(\bz,\blambda,t)=\frac{D(\bz,t)}{2\pi\Sigma^2}\,e^{-\frac{\left|{\blambda-\boldsymbol{\cal A}(\bz)}\right|^2}{2\Sigma^2}}=\frac{D(\bz,t)}{2\pi\Sigma^2}\,e^{-\frac{\left({\lambda_q-p}\right)^2+\lambda_p^2}{2\Sigma^2}}
\,,
\label{WignerClosure}
\eeq
is an exact solution of \eqref{WMeq} provided $D$ satisfies $\partial_tD=\{H,D\}$.
The second equality in \eqref{WignerClosure} follows immediately by recalling $\boldsymbol{\cal A}(\bz)=(p,0)$. The von Neumann operator associated to the Wigner function \eqref{WignerClosure}  is found as
\beq
\widehat{\Theta}(\bz,\bz', t)=\int \!W\Big(\frac{\bz+\bz'}2,\blambda,  t\Big)e^{-i\blambda\cdot(\bz-\bz')}\de^2\lambda= D\Big(\frac{\bz+\bz'}2, t\Big)e^{-\frac1{2\hbar}\big({i}(q-q')(p+p')+\frac{\Sigma^2}{\hbar}|\bz-\bz'|^2\big)}.
\label{VNoper}
\eeq
In the limit case $\Sigma\to0$, the von Neumann operator associated to the  singular Wigner function $W=D\delta(\blambda-\boldsymbol{\cal A})$   was presented explicitly in \cite{GBTr21a}.

Here, the Wigner function \eqref{WignerClosure} appears as a closure for the von Neumann equation \eqref{vneqn}, or its Wigner-Moyal correspondent in \eqref{WMeq}. Actually,  the  function \eqref{WignerClosure} is an exact solution of  equation \eqref{WMeq}, which emerges as the natural extension of KvH theory to Wigner distributions. We notice that, as  mentioned above, the operator \eqref{VNoper} is sign-indefinite and thus is more general than the mixture \eqref{vnoper}. Nonetheless, this unsigned operator yields a positive distribution \eqref{mario}, so that the resulting closure model coincides with the classical Liouville equation for $\rho_c=D$ (or KvN equation, by writing $D=|\chi|^2$).

Equivalently, the  classical Liouville equation may be  obtained by simply replacing $\bsigma\to D\boldsymbol{\cal A}$ in the Euler-Poincar\'e variational principle \eqref{EP-mKvH} for Koopman mixtures. The latter observation allows us to avoid dealing with von Neumann operators when looking for a reduction process taking the KvH construction to the simpler KvN dynamics.

\section{Calculations for the diamond operator\label{Append1}}
From \eqref{funcder}, we compute the diamond $\Upsilon\diamond\delta h/\delta\Upsilon$ term by term as follows

\[
\big\langle\widehat{H}\Upsilon+\boldsymbol{\cal A}_B\cdot\langle\bX_{\widehat{H}}\rangle\Upsilon,\nabla\Upsilon\big\rangle-\big\langle\Upsilon,\nabla(\widehat{H}\Upsilon+\boldsymbol{\cal A}_B\cdot\langle\bX_{\widehat{H}}\rangle\Upsilon)\big\rangle=-\|\Upsilon\|^2\langle\nabla{\widehat{H}}\rangle-\|\Upsilon\|^2\nabla\big(\boldsymbol{\cal A}_B\cdot\langle\bX_{\widehat{H}}\rangle\big)
\]
\begin{align*}\hspace{-1cm}
\big\langle \bX_{\widehat{H}}\cdot(i\hbar\nabla)\Upsilon,\nabla\Upsilon\big\rangle
-
\big\langle\Upsilon,\nabla(\bX_{\widehat{H}}\cdot(i\hbar\nabla)\Upsilon)\big\rangle
=&\ 
2\big\langle \bX_{\widehat{H}}\cdot(i\hbar\nabla)\Upsilon,\nabla\Upsilon\big\rangle-\nabla\big\langle\Upsilon,\bX_{\widehat{H}}\cdot(i\hbar\nabla)\Upsilon\big\rangle
\\
=&\ 
2\big\langle \bX_{\widehat{H}}\cdot(i\hbar\nabla)\Upsilon,\nabla\Upsilon\big\rangle+\nabla(\|\Upsilon\|^2\boldsymbol{\cal A}_B\cdot\langle\bX_{\widehat{H}}\rangle)
\\&
-\nabla\big\langle\Upsilon,(\bX_{\widehat{H}}-\langle\bX_{\widehat{H}}\rangle)\cdot(i\hbar\nabla)\Upsilon\big\rangle
\end{align*}
\begin{align*}\hspace{-1cm}
\big\langle \bX_{\widehat{H}}\cdot\boldsymbol{\cal A}_B\Upsilon,\nabla\Upsilon\big\rangle
-
\big\langle\Upsilon,\nabla(\bX_{\widehat{H}}\cdot\boldsymbol{\cal A}_B\Upsilon)\big\rangle
=&\ 
\big\langle \bX_{\widehat{H}}\cdot\boldsymbol{\cal A}_B\Upsilon,\nabla\Upsilon\big\rangle
-\|\Upsilon\|^2\langle\nabla(\bX_{\widehat{H}}\cdot\boldsymbol{\cal A}_B)\rangle
\\
&\ 
-
\big\langle \Upsilon,\bX_{\widehat{H}}\cdot\boldsymbol{\cal A}_B\nabla\Upsilon\big\rangle
\\
=& \ 
-\|\Upsilon\|^2\langle\nabla(\bX_{\widehat{H}}\cdot\boldsymbol{\cal A}_B)\rangle
\end{align*}
Also if we apply
\begin{align*}
\frac12\bigg(\bigg\langle\frac{\delta h}{\delta \Upsilon},\nabla\Upsilon\bigg\rangle-\bigg\langle\Upsilon,\nabla\frac{\delta h}{\delta \Upsilon}\bigg\rangle\bigg)
=&\,
\bigg\langle\frac{\delta h}{\delta \Upsilon},\nabla\Upsilon\bigg\rangle-\frac12\nabla\bigg\langle\frac{\delta h}{\delta \Upsilon},\Upsilon\bigg\rangle
\,,
\end{align*}
we have
\begin{align*}
\frac12\Big(\Big\langle i\hbar\Upsilon\operatorname{div}\langle\bX_{\widehat{H}}\rangle,\nabla\Upsilon\Big\rangle+\Big\langle i\hbar\Upsilon,\left(\operatorname{div}\langle\bX_{\widehat{H}}\rangle\right)\nabla\Upsilon+\Upsilon\nabla\operatorname{div}\langle\bX_{\widehat{H}}\rangle\Big\rangle\Big)
\\
\hspace{-1.7cm}=&\ 
\|\Upsilon\|^2\boldsymbol{\cal A}_B\operatorname{div}\langle\bX_{\widehat{H}}\rangle
\\
=&\ 
2\boldsymbol{\cal A}_B\big\langle\Upsilon,(\bX_{\widehat{H}}-\langle\bX_{\widehat{H}}\rangle)\cdot\nabla\Upsilon\big\rangle
\end{align*}
and
\[
2\Big\langle i\hbar\langle\bX_{\widehat{H}}\rangle\cdot\nabla\Upsilon,\nabla\Upsilon\Big\rangle-\nabla\Big\langle\Upsilon,\left(i\hbar\langle\bX_{\widehat{H}}\rangle\cdot\nabla\Upsilon\right)\Big\rangle
=
2\big\langle \langle\bX_{\widehat{H}}\rangle\cdot(i\hbar\nabla)\Upsilon,\otimes\nabla\Upsilon\big\rangle+\nabla(\|\Upsilon\|^2\boldsymbol{\cal A}_B\cdot\langle\bX_{\widehat{H}}\rangle)
\,,
\]
which altogether recovers \eqref{vectfield}.

\color{black}
\section{Closure method and hybrid von Neumann operators\label{sec:vNOp}}
In Section \ref{sec:NHWE}, we showed how the relation \eqref{closrel} is used to achieve a positive-definite classical density while taking the Lagrangian \eqref{Tiziana2} into the manifestly gauge-invariant form \eqref{Tiziana3}.
 In this appendix, we will show how the resulting gauge-invariant Lagrangian $\ell_{\scriptscriptstyle QC}$ in \eqref{untangledLagr}  emerges in the context of the original theory from Section \ref{sec:hybrids} by resorting to von Neumann operators and their Wigner functions. 

Following the discussion in Appendix \ref{sec:KoopMixtures}, we prefer to partly abandon the wavefunction picture of Koopman  dynamics in favor of the mixture picture, which was mentioned in \cite{GBTr21} only briefly. The reason for resorting to mixtures is that $\boldsymbol{\cal A}_B$ often vanishes at the initial time (for example, when $\psi$ is real) and thus the relation \eqref{closrel} would require the presence of topological singularities. In more generality, we wish to understand how the hybrid Lagrangian in \eqref{untangledLagr} emerges as a type of closure model by adopting an ansatz, similarly to the arguments in Appendix \ref{sec:KoopMixtures}. In particular, in analogy to \eqref{vnoper}, we combine mixtures with the exact factorization \eqref{EFDef} by constructing a von Neumann operator with matrix elements 
\beq
\widehat{\Theta}(\bz,\bz',x,x')=\sum_{a=1}^Nw_a\Upsilon_a(\bz,x)\Upsilon_a^*(\bz',x')
\,,\qquad\text{with}\qquad
\Upsilon_a(\bz,x)=\chi_a(\bz)\psi(x;\bz)
\,,
\label{vnoper2}
\eeq
and $\int|\psi|^2\,\de x=1$. Mixtures of this type are analogous to those considered in \cite{FoHoTr19} within the context of nonadiabatic quantum hydrodynamics. Then,  following the same arguments as in Appendix \ref{sec:KoopMixtures}, we will proceed by inserting the  exact factorization $\Upsilon_a=\chi_a\psi$ in the  mixture Lagrangian $\sum_aw_aL_{\rm QC}(\Upsilon_a,\partial_t\Upsilon_a)$, where $L_{\rm QC}$ is given in \eqref{VP-Hyb1}. After writing $\chi_a=\sqrt{D_a}e^{iS_a/\hbar}$ and using the definitions \eqref{momaps}, the steps from  Section \ref{sec:EF} take us to the following Euler-Poincar\'e Lagrangian  (and thus to its associated dynamics):
\beq
L=\int \!\Big(\bsigma\cdot\boldsymbol{\cal X}+D\big\langle\psi,i\hbar\partial_t\psi-{\widehat{H}}\psi+i\hbar\widetilde{\bX}_{\widehat{H}}\cdot\nabla\psi-(D^{-1}\bsigma+\boldsymbol{\cal A}_B-\boldsymbol{\cal A})\cdot\bX_{\widehat{H}}\psi\big\rangle\Big)\,\de^2z
\,.
\label{Tiziana2bis}
\eeq 
In this case, the original expression \eqref{densities2} of the classical density changes to $\rho_c=D+\operatorname{div}(J\bsigma+DJ\boldsymbol{\cal A}_B-DJ\boldsymbol{\cal A})$ and one would like to set $D^{-1}\bsigma+\boldsymbol{\cal A}_B=\boldsymbol{\cal A}$ in such a way to obtain a positive-definite  density $\rho_c=D$. Unlike the purely classical case treated in Appendix \ref{sec:KoopMixtures}, the hybrid relation $D^{-1}\bsigma+\boldsymbol{\cal A}_B=\boldsymbol{\cal A}$   is not preserved in time by the  dynamics arising from \eqref{Tiziana2bis}. Here, we will devise  a closure scheme so that a particular  ansatz at the level of the von Neumann operator is applied to the Lagrangian \eqref{Tiziana2bis} thereby returning  \eqref{Tiziana3}.

We realize that the  operator \eqref{vnoper2} emerges as a special case of a wider class of operators, which will be particularly convenient in designing our closure scheme. Specifically, let us consider matrix elements   of the type
\beq
\widehat{\Theta}(\bz,\bz',x,x')=\hat\theta(\bz,\bz')\psi(x;\bz)\psi^*(x';\bz')
\,,
\label{closVNOp}
\eeq
where $\hat\theta$ is itself a von Neumann operator acting only on the classical sector.  We observe that \eqref{closVNOp} leads to the relations \eqref{momaps2} with $\widehat\Theta$ replaced by $\hat\theta$. In this setting, the Hamiltonian functional $h(\hat\theta,\psi)$ is given by \eqref{hamilt1} upon replacing $D\nabla S\to\bsigma$ and using  \eqref{momaps2} to express $D$ and $\bsigma$ in terms of $\hat\theta$. Thus, if we follow the procedure in Appendix \ref{sec:KoopMixtures} and define $\bLambda:=D^{-1}\bsigma$,  finding a closure for which $\bLambda+\boldsymbol{\cal A}_B=\boldsymbol{\cal A}$ amounts to finding a von Neumann operator $\hat\theta$ so that $\big[\hat{\blambda},\hat\theta\big]_+(\bz,\bz')|_{\bz'=\bz}=2\hat \theta  (\bz,\bz')|_{\bz'=\bz}(\boldsymbol{\cal A}-\boldsymbol{\cal A}_B)$.
 We will address this problem by resorting to Wigner transforms, following the treatment in Appendix \ref{sec:KoopMixtures}. We anticipate that, similarly to the approach in that section, we will now allow for the operator $\hat\theta$ to be sign-indefinite, as long as its associated phase-space distribution $\rho_c=D+\operatorname{div}(J(\bsigma+D\boldsymbol{\cal A}_B-D\boldsymbol{\cal A}))$ remains positive.

Let us introduce the Wigner transform 
\beq
w(\bz,\blambda)=\frac1{(\hbar\pi)^2}\int\!\hat\theta(\bz-\bz',\bz+\bz')\,e^{2i\blambda\cdot\bz'/\hbar}\,\de^2z'
\label{wignclos}
\eeq
of the von Neumann operator $\hat\theta$, so that
\beq\label{momaps3}
D(\bz)=\int\! w(\bz,\blambda)\,\de^2\lambda
\,,\qquad\qquad
D(\bz)\bLambda(\bz)=\int \!\blambda w(\bz,\blambda)\,\de^2\lambda
\,.
\eeq
Then, we see that the relation $\bLambda+\boldsymbol{\cal A}_B=\boldsymbol{\cal A}$ is achieved by 
a slight modification of the closure \eqref{WignerClosure}, that is
\beq
w(\bz,\blambda)=\frac{D(\bz)}{2\pi\Sigma^2}\,\exp\!\bigg(\!-\frac{\left|{\blambda-\boldsymbol{\cal A}(\bz)+\boldsymbol{\cal A}_B(\bz)}\right|^2}{2\Sigma^2}\bigg)
\,.
\label{WignerClosure2}
\eeq
Here, we recall that the Berry connection $\boldsymbol{\cal A}_B$ is expressed in terms of $\psi$ by \eqref{berryconn}. 
Thus, we are left with the conclusion that the closure relation 
$
\bLambda=\boldsymbol{\cal A}-\boldsymbol{\cal A}_B
$
taking the Lagrangian \eqref{Tiziana2} into the closure model Lagrangian \eqref{Tiziana3} 
is achieved by a von Neumann operator of the type \eqref{closVNOp}, where $\hat\theta$ has a Wigner transform given by \eqref{WignerClosure2} for some $\Sigma$. Similarly to \eqref{VNoper}, we compute
\beq
\hat{\theta}(\bz,\bz')=D(\bzeta)\exp\!\Big(-\frac{i}{2\hbar}(\bz-\bz')\cdot(\boldsymbol{\cal A}(\bzeta)-\boldsymbol{\cal A}_B(\bzeta))-\frac{\Sigma^2}{2\hbar^2}|\bz-\bz'|^2\Big)\Big|_{\bzeta=(\bz+\bz')/2}
\,,
\label{VNoper2}
\eeq
where we recall $\boldsymbol{\cal A}(\bz)=(p,0)$. In the limit $\Sigma\to 0$, this type of von Neumann  operator was previously found in \cite{FoHoTr19}, although the expression \eqref{VNoper2} now involves phase-space coordinates. By following analogous steps to those in \cite{FoHoTr19} (see Section 5.1 therein), one shows that replacing \eqref{VNoper2} in the variational principle underlying the dynamics of the hybrid von Neumann operator in \eqref{closVNOp} ultimately leads to the  Lagrangian \eqref{untangledLagr}, thereby recovering the NHWE system \eqref{NHWE}-\eqref{vectfield}. The explicit relations involved in this argument are omitted here and we refer to  Section 5.1  in \cite{FoHoTr19} for the technical details  in the fully quantum context.

We notice that the Wigner function  \eqref{WignerClosure2} extends the usual fluid closure method from kinetic theory to consider Maxwellian distributions on phase-space. Although their von Neumann operator is sign-indefinite, as we pointed out in Appendix \ref{sec:KoopMixtures}, Wigner distributions of general Maxwellian type are occasionally used in mixed-state quantum hydrodynamics \cite{HuBu12,LiHaHe89}. In the present context, however, we will not have to deal with the various aspects concerning Maxwellian distributions. Indeed, as we already showed, our NHWE model is simply obtained by replacing $\nabla S\to\boldsymbol{\cal A}-\boldsymbol{\cal A}_B$ in the Lagrangian \eqref{Tiziana2} from the original theory.
\color{black}

\end{document}